\documentclass[a4paper]{paper_cw}

\usepackage{helvet}
\usepackage{amsmath}
\usepackage{amssymb}
\usepackage{mathrsfs}
\usepackage{graphics}
\usepackage{natbib}
\usepackage{color}
\usepackage{fancyhdr} 
\usepackage{colortbl} 
\usepackage{subfigure}
\usepackage[dvips]{graphicx} 
\usepackage{marvosym}
\usepackage{natbib}
\usepackage{import}

\long\def\symbolfootnote[#1]#2{\begingroup \def\thefootnote{\fnsymbol{footnote}}\footnote[#1]{#2} \endgroup} 

\begin{document}

\title{Optimal control of dielectric elastomer actuated multibody dynamical systems}

\author{Dengpeng Huang \and Sigrid Leyendecker}

\institute{D. Huang(\Letter), S. Leyendecker  \at Institute of Applied Dynamics, Friedrich-Alexander-Universität Erlangen-Nürnberg, Germany
\\\email{dengpeng.huang@fau.de}}

\maketitle
\thispagestyle{empty}

\abstract{In this work, a simulation model for the optimal control of dielectric elastomer actuated flexible multibody dynamics systems is presented. The Dielectric Elastomer Actuator (DEA) behaves like a flexible artificial muscles in soft robotics. It is modeled as an electromechanically coupled geometrically exact beam, where the electric charges serve as control variables. The DEA-beam is integrated as an actuator into multibody systems consisting of rigid and flexible components. The model also represents contact interaction via unilateral constraints between the beam actuator and e.g.~a rigid body during the grasping process of a soft robot. Specifically for the DEA, a work conjugated electric displacement and strain-like electric variables are derived for the Cosserat beam. With a mathematically concise and physically representative formulation, a reduced free energy function is developed for the electromechanically coupled beam. In the optimal control problem, an objective function is minimized while the electromechanically coupled dynamic balance equations for the multibody system have to be fulfilled together with the complementarity conditions for the contact and boundary conditions. The optimal control problem is solved via a direct transcription method, transforming it into a constrained nonlinear optimization problem. The electromechanically coupled geometrically exact beam is firstly semidiscretized with 1D finite elements and then the multibody dynamics is temporally discretized with a variational integrator leading to the discrete Euler-Lagrange equations, which are further reduced with the null space projection. The discrete Euler-Lagrange equations and the boundary conditions serve as equality constraints, whereas the contact constraints are treated as inequality constraints in the optimization of the discretized objective. The constrained optimization problem is solved using the IPOPT solver. The effectiveness of the developed model is demonstrated by three numerical examples, including a cantilever beam, a soft robotic worm and a soft grasper.}


\section{Introduction}
Due to superior performances such as a higher efficiency in energy, completing complex tasks and a safe interaction with environment, soft robotics are highly required in industrial production, medical treatment and daily life. The movement and action of soft robotics rely on soft actuators inserted in flexible multibody systems. Dielectric Elastomer Actuators (DEAs) have often been selected as artificial muscles for soft robotics, see \citep{bar2000electroactive}, \citep{lowe2005dielectric}, \citep{kovacs2009stacked} and \citep{duduta2019realizing} for instance. As the basic element of DEA, the capacitor is made by sandwiching a dielectric elastomer between two compliant electrodes. When external electric potentials are applied to the compliant electrodes, the dielectric elastomer is polarized, resulting in an electrostatic pressure, see the electromechanically coupled models in \citep{pelrine1998electrostriction}, \citep{wissler2007electromechanical}, \citep{suo2008nonlinear} and \citep{schlogl2017polarisation} for instance. The electrostatic pressure serves as the driving forces in the elastomer to generate deformations for actuation. To obtain larger deformations, multiple capacitors are stacked on top of each other in real applications. 

The numerical simulation of the DEA is based on the formulations of nonlinear electroelasticity. The fundamental theory of nonlinear electroelasticity has been presented in \citep{1978meto} concerning the interaction of electromagnetic and elastic fields in deformable continua, in \citep{dorfmann2005nonlinear} for finite deformations and in \citep{vu2007numerical} formulating variational electro- and magneto-elastostatics. The electromechanically coupled free energy function for dielectric elastomers is reported by \citep{zhao2007electromechanical}, \citep{vu2007numerical} and \citep{suo2010theory}. A viscoelastic 3D finite element model of the DEA is developed by \citep{schlogl2016electrostatic} for the dynamic analysis. 

The 3D finite element model of the DEA is extended to flexible multibody dynamics system in \citep{schlogl2016dynamic}. However, the high computational cost induced by huge amount of degrees of freedom in large 3D finite element models leads to difficulties for the advanced simulations of DEAs, such as in flexible multibody dynamics simulation and their optimal control. To this end, the electromechanically coupled beam dynamics model is developed with a variational integrator scheme by \citep{huang2022electromechanically}, where several advantages compared to the continuum model can be observed, such as less degrees of freedom by still representing all deformation modes and a good long term energy behaviour. However, the extension to DEA actuated flexible multibody system dynamics is still missing. The formulation of flexible multibody system dynamics can be found e.g.~in \citep{leyendecker2008discrete}. The time integration of constrained geometrically nonlinear beam dynamics has been discussed by \citep{ARMERO20012603} considering energy conserving/decaying algorithms, in \citep{betsch2006discrete} with energy-momentum schemes with the null space method and in \citep{leyendecker2008variational} using the variational integrators.

Apart from the forward dynamics simulation, the optimal control of the DEA is a key part for its applications. For instance, the electric potential applied to the DEAs of soft robotics has to be determined and controlled for every motion trajectory of a robotic mechanism to perform an action between an initial and final state. This kind of problem can be formulated mathematically by a constrained optimisation problem, also known as the optimal control problem, where the objective function is defined according to the specific physical situation. In the context of optimal control theory, the Discrete Mechanics and Optimal Control (DMOC) method is developed in \citep{ober2011discrete} by combining a direct transcription optimal control method with structure preserving time stepping. Further, the Discrete Mechanics and Optimal Control for Constrained systems (DMOCC) is developed in \citep{leyendecker2010discrete} to account for holonomic constraints in the formulation of multibody dynamics. Other applications of DMOC can be found in the context of non-holonomically constrained robotics \citep{kobilarov2007optimal}, two-dimensional compass biped gait \citep{leyendecker2013structure} and space mission design \citep{moore2009optimization} for instance. In the context of inverse dynamics, the servo-constraints have been applied in the feedforward control of underactuated mechanical systems, such as the cranes in \citep{otto2018real}. In \citep{strohle2022simultaneous}, the servo-constraints problem is extended to finite dimensional systems by using a simultaneous space-time discretization approach. However, the control problem in the electromechanically coupled flexible multibody system dynamics has not been addressed.

The objective of this work is to formulate and solve the optimal control problem of the DEA actuated flexible multibody system dynamics, where the electromechanically coupled beam represents the DEA and the electric charges on the beam serve as control variables. Complementarity conditions represent the contact between the DEA-beam actuators and e.g. rigid bodies in a grasping process.

After direct transcripton, the forward dynamics equations of the electromechanically coupled flexible multibody system are formulated as the equality constraints. The contact between the beam and the rigid body is considered in the inequality constraints, by which the grasping with soft robotic fingers can be simulated. Instead of using the free energy function formulated for the 3D continuum, a mathematically concise and physically representative free energy function is formulated for the electromechanically coupled beam.

This paper is structured as follows: In Section 2, the kinematics, the work-conjugated variables and a reduced free energy function for the electromechanically coupled beam are presented. Then the variational formulation of the electromechanically coupled flexible multibody system dynamics is derived in the continuous and the discretized setting in Section 3. Section 4 presents the direct transcription of the optimal control problem for the electromechanically coupled multibody dynamics system. Numerical examples are presented in Section 5, followed by the conclusions in Section 6.

\section{Electromechanical coupling in geometrically exact beam}
The governing equations for electromechanical coupling in the continuum as well as the beam have been introduced in \citep{huang2022electromechanically}. In this section, the mechanical and electrical kinematics of the geometrically exact beam is firstly summarized. Then, the electric displacements and the strain-like electrical variables for the beam are derived from the {work conjugated variables} in the continuum. To build a mathematically concise {and physically representative} {free energy function} for the beam, the {free energy} of the continuum is reduced to obtain a beam-specific {free energy function}.
\subsection{Kinematics}
\begin{figure}[htb!]
	\centering
	\includegraphics[width=.6\textwidth]{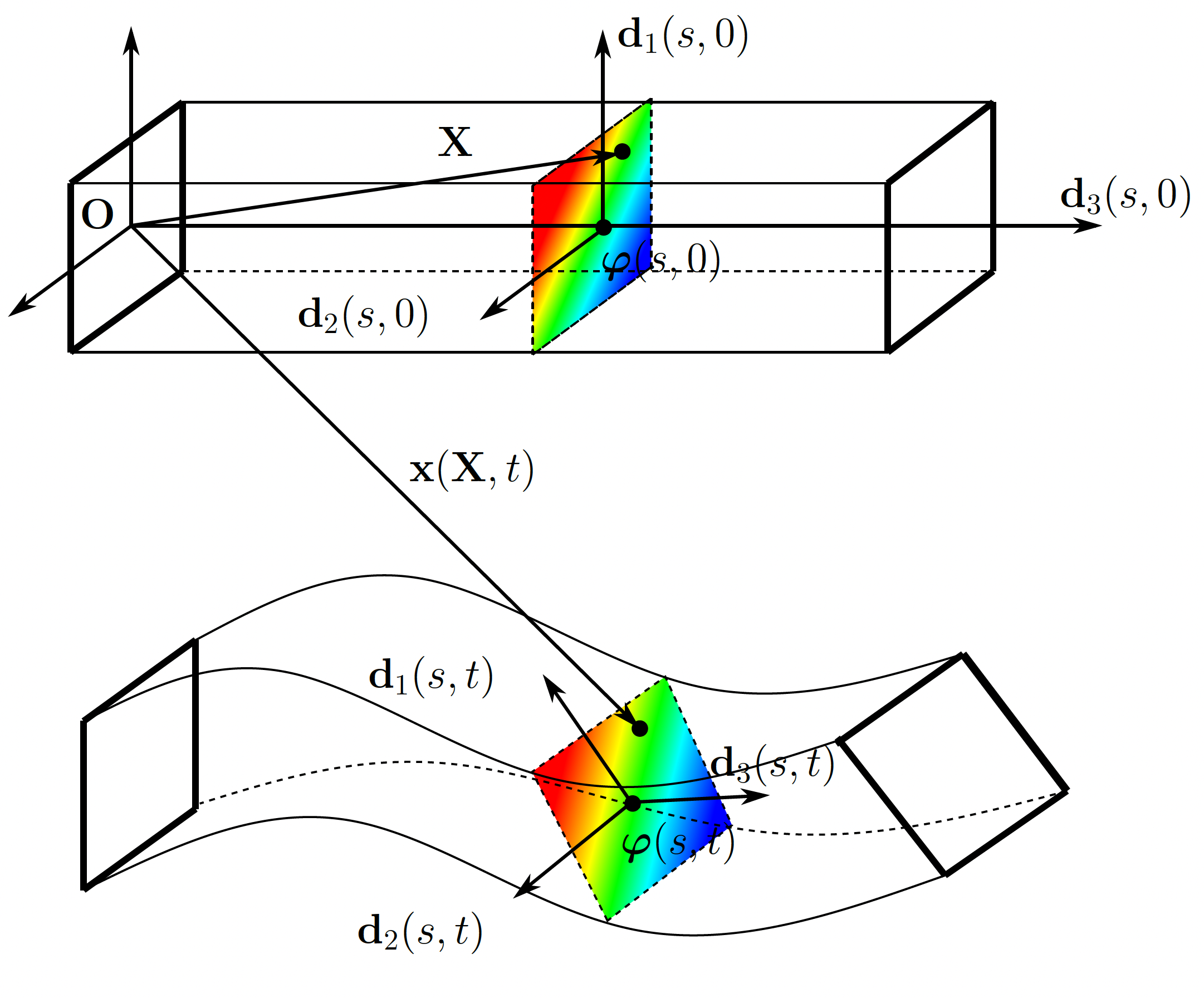}
	\caption{Configurations of the beam.}
	 \label{config}
\end{figure}

The deformation state of an initially straight beam over time can be {described} by the initial configuration and the current configuration, as shown in Fig.~\ref{config}.

The geometrically exact beam is represented as a special Cosserat continuum, see \citep{antmannonlinear}, where the beam is represented with a 3D curve in its longitude direction and the cross sections along the beam curve are assumed to be rigid. Apart from the centroids $\boldsymbol{\varphi}\in \mathbb{R}^3$ on the curve, the orthonormal triads $\mathbf{d}_i  (i=1,2,3)$ are attached to the beam centroids to characterize the orientation of the rigid cross sections. Th placement of a material point in the current configuration of the beam is given by
\begin{align}
\mathbf{x}(X^k,s,t)=\boldsymbol{\varphi}(s,t)+X^k \mathbf{d}_k(s,t), \;\;\;\; k=1,2. \label{x}
\end{align}
where $s \in [0,L] \subset \mathbb{R}$ denotes the arc-length of the line of the centroids $\boldsymbol{\varphi}(s,0)\in \mathbb{R}^3$ in the initial configuration, the directors $\mathbf{d}_k(s,t), k=1,2$ span a principle basis of the cross section, $X^k$ are the curvilinear coordinates on {the} cross section and {$t$ is time}.

The electrical kinematics are formulated for the Cosserat beam in \citep{huang2022electromechanically}, where the electric potential at the point $(X^1,X^2,s)$ on the beam cross section is written in a similar form as Eq.~(\ref{x})
\begin{align} 
	\phi (X^k, s, t)=\phi _o(s, t) + X^1 \alpha(s, t) + X^2 \beta(s, t)
	\label{linearV}
\end{align}
with $\phi _o(s, t) $ the electric potential at the beam node, $ \alpha(s, t)$ and $\beta(s, t)$ the incremental parameters of the electric potential in the directions of $\mathbf{d}_1(s, t)$ and $\mathbf{d}_2(s, t)$, respectively. Eq.~(\ref{linearV}) provides a linear distribution of the electric potential on the cross section, which is the minimum requirement for actuating complex beam deformations. The extension to nonlinear distributions of the electric potential on beam cross sections can be made by adding higher order terms.

By setting the origin of the global Cartesian coordinate system $\mathbf{O}$ to one end of the beam, the deformation gradient at a point $(X^1, X^2, s)$ in the beam can be written as, see \citep{auricchio2008geometrically},
\begin{align}
\mathbf{F}(X^k, s)&=\frac{\partial \mathbf{x}}{\partial \mathbf{X}}=\frac{\partial \mathbf{x}}{\partial X_i} \otimes  \mathbf{d}_i(s,0) \nonumber\\
&=\left[ \mathbf{I} + \left(\frac{\partial \boldsymbol{\varphi}(s,t)}{\partial s} - \mathbf{d}_3(s,t) + X^1 \frac{\partial \mathbf{d}_1(s,t)}{\partial s} + X^2 \frac{\partial \mathbf{d}_2(s,t)}{\partial s} \right)\otimes  \mathbf{d}_3(s,t) \right] \boldsymbol{\Lambda}(s),
\label{F}
\end{align}
with the rotation tensor $ \boldsymbol{\Lambda}(s)=\mathbf{d}_i(s,t) \otimes  \mathbf{d}_i(s,0)$ and $\boldsymbol{\Lambda}(s)^{-1}=\boldsymbol{\Lambda}(s)^T $. 

Correspondingly, the electric field at the point $(X^1, X^2, s)$ is defined as the gradient of the electric potential in Eq.~(\ref{linearV})
\begin{align} 
\mathbf{E}^e(X^k,s)&= -\frac{\partial \phi}{\partial X_i} \otimes  \mathbf{d}_i(s,0)\\
&=-\left[ \alpha(s, t)  \mathbf{d}_1(s,0) + \beta(s, t)  \mathbf{d}_2(s,0) + \left( \frac{\partial \phi_o(s, t)}{\partial s}  + X^1 \frac{\partial  \alpha(s, t)}{\partial s} + X^2 \frac{\partial  \beta(s, t)}{\partial s} \right) \mathbf{d}_3(s,0) \right]. \label{Ee}
\end{align}
It can be observed that $\mathbf{F}(X^k, s)$ and $\mathbf{E}(X^k, s)$ are not objective since they depend on the directors. Instead of the deformation gradient and the electric field mentioned above, the strain of the geometrically exact beam is measured by the beam-specific kinematic variables such as the beam strains in the current configuration {in the} global basis
\begin{align}
\boldsymbol{\gamma} = \boldsymbol{\varphi}_{,s}(s,t) - \mathbf{d}_3(s,t), \qquad \boldsymbol{\kappa} =  \frac{1}{2} \mathbf{d}_i(s,t) \times \mathbf{d}_{i,s}(s,t),
\end{align}
and {in the} local basis
\begin{align}
 \boldsymbol{\gamma}&=\gamma_i \mathbf{d}_i(s,t),\qquad  \gamma_i=\boldsymbol{\gamma}\cdot \mathbf{d}_i(s,t)=\boldsymbol{\varphi}_{,s}(s,t) \cdot \mathbf{d}_i(s,t)-\delta_{i3}, \nonumber\\
  \qquad \boldsymbol{\kappa}&=\kappa_i \mathbf{d}_i(s,t),\qquad  \kappa_j =\boldsymbol{\kappa}\cdot \mathbf{d}_j(s,t)= \frac{1}{2} \epsilon_{jkl} \mathbf{d}_{k,s}(s,t) \cdot \mathbf{d}_l(s,t), \label{strain}
\end{align}
where $\boldsymbol{\gamma}$ is the beam strain for shear and elongation, $\boldsymbol{\kappa}$ is the beam strain for torsion and bending, $\delta_{ij}$ is the Kronecker delta and $ \epsilon_{jkl}$ is the permutation symbol. The strains in the current configuration can be transformed to the reference configuration by
\begin{align}
	\boldsymbol{\Gamma}=\boldsymbol\Lambda^T \boldsymbol{\gamma}, \qquad\mathbf{K}=\boldsymbol\Lambda^T \boldsymbol{\kappa}.
\end{align}
The beam strains in Eq.~(\ref{strain}) are usually used in the beam {free energy function} since the strain components in the local basis are independent of the local frame (objective), i.e.
\begin{align}
\gamma_i =\boldsymbol{\Gamma}\cdot \mathbf{d}_i(s,0)=\boldsymbol{\gamma}\cdot \mathbf{d}_i(s,t), \qquad \kappa_i=\mathbf{K}\cdot \mathbf{d}_i(s,0)=\boldsymbol{\kappa}\cdot \mathbf{d}_i(s,t).
\end{align}	
Corresponding to the mechanical strain measures, the strain-like electrical variables are required to model the electromechanical coupling in the beam, which will be derived through the {work conjugated variables} in the next section.

\subsection{Work-conjugated variables}
For the later computing of the virtual work induced by the electric field, the variation of the electric field (\ref{Ee}) is written as
\begin{align} 
\delta \mathbf{E}^e = -\left[ \delta \alpha(s,t)  \mathbf{d}_1(s,0) + \delta \beta(s,t)  \mathbf{d}_2(s,0) + \delta\left( \frac{\partial \phi_o(s,t)}{\partial s}  + X^1 \frac{\partial  \alpha(s,t)}{\partial s} + X^2 \frac{\partial  \beta(s,t)}{\partial s} \right) \mathbf{d}_3(s,0) \right]. \label{dE}
\end{align}
In continuum electomechanics, the electric field $\mathbf{E}^e$ is work-conjugated with the electric displacement $\mathbf{D}$, where the internal virtual work $\delta W_{\rm int}$ induced by the electric field is given by
\begin{align} 
\delta W_{\rm int}=\int_{B_0} \mathbf{D}\cdot \delta \mathbf{E}^e dV.
\end{align}
By applying the electric displacement and the variation of the electric field in Eq.~(\ref{dE}) to the internal virtual work, we obtain
\begin{align} 
	\delta W_{\rm int} = & \int_{B_0} -(D_i \mathbf{d}_i^0) \cdot  \left[ \delta \alpha(s,t)  \mathbf{d}_1^0 + \delta \beta(s,t)  \mathbf{d}_2^0 + \delta\left( \frac{\partial \phi_o(s,t)}{\partial s}  + X^1 \frac{\partial  \alpha(s,t)}{\partial s} + X^2 \frac{\partial  \beta(s,t)}{\partial s} \right) \mathbf{d}_3^0 \right]  dV\\
	=& \int_c  -\left[ \delta \alpha(s,t) \int_{\Sigma} D_1 dA +\delta \beta(s,t)  \int_{\Sigma} D_2 dA + \delta \frac{\partial \phi_o(s,t)}{\partial s} \int_{\Sigma} D_3 dA \right. \nonumber\\ 
  &	\left. + \delta \frac{\partial \alpha(s,t)}{\partial s} \int_{\Sigma} X^1 D_3 dA + \delta \frac{\partial \beta(s,t)}{\partial s} \int_{\Sigma} X^2 D_3 dA \right]ds \\
	=&\int_c  \mathbf{D}_{\Xi} \cdot \delta \boldsymbol{\Xi}   ds + \int_c  \mathbf{D}_{\Theta} \cdot \delta \boldsymbol{\Theta} ds,
\end{align}
where {$\mathbf{d}_i^0=\mathbf{d}_i(s,0)$, $c$ denotes the beam curve, $\Sigma$ denotes the beam cross section,} $(\mathbf{D}_{\Xi} , \boldsymbol{\Xi})$ and $(\mathbf{D}_{\Theta}, \boldsymbol{\Theta})$ are two pairs of the conjugated electric displacement and strain-like electrical variable formulated as
\begin{align}
	\mathbf{D}_{\Xi} &= \int_{\Sigma} D_1 dA \mathbf{d}_1^0+ \int_{\Sigma} D_2 dA\mathbf{d}_2^0+ \int_{\Sigma} D_3 dA \mathbf{d}_3^0,  \qquad
	\boldsymbol{\Xi}  = -\alpha\mathbf{d}_1^0 -\beta\mathbf{d}_2^0 -\frac{\partial \phi_o }{\partial s} \mathbf{d}_3^0, \\
	\mathbf{D}_{\Theta} &=  \int_{\Sigma} X^1 D_3 dA\mathbf{d}_1^0+ \int_{\Sigma} X^2 D_3 dA \mathbf{d}_2^0+ 0 \mathbf{d}_3^0, \qquad
	\boldsymbol{\Theta}  = -\frac{\partial \alpha }{\partial s} \mathbf{d}_1^0 -\frac{\partial \beta }{\partial s}\mathbf{d}_2^0+ 0\mathbf{d}_3^0.
\end{align}
The electric displacements $(\mathbf{D}_{\Xi}, \mathbf{D}_{\Theta})$ and the strain-like electrical variables $(\boldsymbol{\Xi}, \boldsymbol{\Theta})$ are formulated with the reference local basis. The reference strain-like electrical variables can be transformed into the current configuration by
\begin{align}
	\boldsymbol{\xi}=\boldsymbol\Lambda \boldsymbol{\Xi}, \qquad\boldsymbol{\theta}=\boldsymbol\Lambda \boldsymbol{\Theta}.
\end{align}

\subsection{A reduced {free energy function}}
The electromechanically coupled {free energy} for the beam has been derived consistently from continuum electromechanics in \citep{huang2022electromechanically}, which leads to the {corresponding transfer} of material models in the continuum to the beam. However, the formulation of the beam {free energy function} is not as mathematically concise as the widely used Saint-Venant-Kirchhoff model \citep{simo1986three} for the hyperelastic beam. To obtain a physically representative and mathematically concise formulation, a reduced {free energy function} is derived in this work to describe the material behavior of the dielectric elastomer beam.

The reduced material model for the beam is developed by rewriting of the {free energy function} in the continuum. In \citep{schlogl2016electrostatic}, the {free energy} density of the continuum is given by
\begin{align}
\Omega(\mathbf{C},\mathbf{E}^e)=\underbrace{ \frac{\mu}{2} \left( \mathbf{C}:\mathbf{1}-3 \right) - \mu {\rm ln} J + \frac{\lambda}{2} ({\rm ln} J)^2}_{\text{Neo-Hookean}} +\underbrace{c_1 \mathbf{E}^e \cdot \mathbf{E}^e + c_2 \mathbf{C}:(\mathbf{E}^e \otimes \mathbf{E}^e)}_{\text{polarization in dielectric material}} - \underbrace{ \frac{1}{2} \varepsilon_0 J \mathbf{C}^{-1} : (\mathbf{E}^e \otimes \mathbf{E}^e)}_{\text{free space term in vacuum}} \label{omega}
\end{align}
with $\mathbf{C}$ the right Cauchy Green tensor, $\lambda$ and $\mu$ the Lam{\'e} parameters, $\varepsilon_0$ the vacuum permittivity, $c_1$ and $c_2$ the electrical parameters, and $J={\rm det}(\mathbf{F})$. The {free energy density} is composed of three parts, the Neo-Hookean part referring to the pure elastic behavior, the polarization part referring to the polarization in the condensed matter and the free space part referring to the effect in vacuum. The electromechanical coupling is characterized by the last two terms in Eq.~(\ref{omega}). To develop the reduced model, the Neo-Hookean part in Eq.~(\ref{omega}) is replaced with the widely used Saint-Venant-Kirchhoff model, whereas the electromechanically coupled part is obtained by {using} the linear part of the right Cauchy Green tensor in the polarization term
\begin{align}
	\Omega^r_e(\boldsymbol{\Gamma}, \mathbf{K}, \boldsymbol{\Xi}, \boldsymbol{\Theta})=c_1 \mathbf{E}^e \cdot \mathbf{E}^e + c_2 \mathbf{C}^{\rm linear}:(\mathbf{E}^e \otimes \mathbf{E}^e), \label{redu}
\end{align}
where the linear part of the right Cauchy Green tensor $\mathbf{C}^{\rm linear}$ is obtained by taking the first two terms of $\mathbf{C}=\mathbf{I}+2(\mathbf{a}^r \otimes \mathbf{d}^0_3)^{\rm sym}+(\mathbf{a}^r\cdot \mathbf{a}^r)\mathbf{d}^0_3 \otimes \mathbf{d}^0_3$ with $\mathbf{a}^r=\boldsymbol{\Gamma}+\mathbf{K} \times X_k\mathbf{d}_k^0$. Since the coefficient $\varepsilon_0$ has a very small value of $8.854 \times 10^{-12}C/Vm$, the free space term in Eq.~(\ref{omega}) is neglected in this work. The details for deriving the reduced model in Eq.~(\ref{redu}) can be found in the Appendix.

 The electromechanically coupled {free energy density} for the beam is obtained by integrating $\Omega^r_e$ over the cross section
\begin{align}
	\Omega^b_e(\boldsymbol{\Gamma}, \mathbf{K}, \boldsymbol{\Xi}, \boldsymbol{\Theta}) =\int_{\Sigma} \Omega^r_e(\boldsymbol{\Gamma}, \mathbf{K}, \boldsymbol{\Xi}, \boldsymbol{\Theta})dA.
\end{align}
By combining the widely used Saint-Venant-Kirchhoff model and the above coupled term, the reduced {free energy density} for the beam is given by
\begin{align}
	\Omega^b &(\boldsymbol{\Gamma}, \mathbf{K}, \boldsymbol{\Xi}, \boldsymbol{\Theta}) = \underbrace{\frac{1}{2} \boldsymbol{\Gamma}^T \mathbf{D}_1 \boldsymbol{\Gamma} + \frac{1}{2} \mathbf{K}^T \mathbf{D}_2 \mathbf{K}}_{\rm mechanical} + \underbrace{(c_1+c_2)\boldsymbol{\Xi}^T \mathbf{D}_3\boldsymbol{\Xi} +(c_1+c_2)\boldsymbol{\Xi}^T \mathbf{D}_4 \boldsymbol{\Theta}}_{\rm electrical} \nonumber\\ 
	&+ \underbrace{2c_2 \Xi_3 \boldsymbol{\Xi}^T \boldsymbol{\Gamma} + 2c_2\Gamma_3( \Theta_1^2 I_1 +\Theta_2^2 I_2)}_{\rm \boldsymbol\Xi, \boldsymbol\Theta \, coupled \, with \, \boldsymbol\Gamma} + \underbrace{2c_2 K_3(\Xi_2\Theta_1 I_1 -\Xi_1\Theta_2 I_2) + 4c_2\Xi_3( \Theta_2K_1I_2 - \Theta_1K_2I_1)}_{\rm \boldsymbol\Xi, \boldsymbol\Theta \, coupled \, with \, \mathbf{K}}, \label{Ob}
\end{align}
where the material tangents are given by
\begin{align}
	\mathbf{D}_1&={\rm diag}\left\lbrace GA, GA, EA \right\rbrace,\qquad \mathbf{D}_2={\rm diag}\left\lbrace EI_1, EI_2, GJ \right\rbrace, \nonumber\\
	\mathbf{D}_3&=(c_1+c_2)A\mathbf{I}, \qquad
	\mathbf{D}_4={\rm diag}\left\lbrace (c_1+c_2)I_1, (c_1+c_2)I_2, 0 \right\rbrace
\end{align}
with $I_k=\int_{\Sigma} X_k^2 dA, k=1,2$ being the second moments of area of the cross section, $J=I_1+I_2$ being the polar moment of area, $A$ being the area of cross section, $E$ and $G$ being the Young's modulus and the shear modulus, respectively. It can be observed that $\Omega^b$ is objective. {Note that} the pure shear and the pure bending can not be actuated by the electric field via this material model, because the shear strain components ($\Gamma_1,\Gamma_2$) and the bending strain components ($K_1,K_2$) are coupled via the electrical component $\Xi_3$ for contraction in the {free energy function}, respectively. This behavior is consistent with the {continuum} model as well as the physics of the dielectric elastomer.

\section{Electomechanically coupled multibody system dynamics}
\subsection{Variational formulation in continuous setting}
The variational formulation of the electomechanically coupled multibody dynamics problem is based on the Lagrange-d'Alembert principle. For a system formulation in redundant coordinates subject to holonomic constraints, the Lagrange-d'Alembert principle reads, see \citep{leyendecker2008variational} and \citep{leyendecker2010discrete},
\begin{align}
\delta \int_{0}^{T} \left[  L( \mathbf{q}, \dot{\mathbf{q}}) - \mathbf{g}^T(\mathbf{q})\cdot \boldsymbol\lambda  - \mathbf{g}^{cT}(\mathbf{q})\cdot \boldsymbol\lambda^c \right] dt + \int_{0}^{T}\mathbf{f}^{\rm ext}( t) \cdot \delta \mathbf{q}dt=0, \label{dAp2}
\end{align}
where  $\mathbf{q}$ is the configuration, $L( \mathbf{q}, \dot{\mathbf{q}})$ is the Lagrangian, $\mathbf{g}$ represents holonomic constraints, $\mathbf{g}^c$ represents contact constraints, $\boldsymbol\lambda$ and $\boldsymbol\lambda^c$ are the Lagrangian multipliers, and $\mathbf{f}^{\rm ext}(t)$ is the external source term containing forces and electric charges. When considering the electrical effect in a geometrically exact beam, the electric potential $\phi_o$ and the incremental variables $(\alpha, \beta )$ in Eq.~(\ref{linearV}) are treated as the electrical degrees of freedom $\boldsymbol\phi=\begin{bmatrix}  \phi_o& \alpha& \beta \end{bmatrix}$. {Similar to \citep{schlogl2016dynamic}}, the configuration of the electromechanically coupled beam model reads
\begin{align}
\mathbf{q}=\begin{bmatrix}
\boldsymbol{\varphi}& \mathbf{d}_1
&\mathbf{d}_2&\mathbf{d}_3& \boldsymbol\phi
\end{bmatrix}^T. \label{q}
\end{align}

The continuous Lagrangian contains the difference between the kinetic energy $T(\dot{\mathbf{q}}) $ and the internal potential energy $V(\mathbf{q})$
\begin{align}
L(\mathbf{q}, \dot{\mathbf{q}})= T(\dot{\mathbf{q}}) -V(\mathbf{q}).
\end{align}
Since the electrical variables do not contribute to the kinetic energy, the kinetic energy for the geometrically exact beams is computed as
\begin{align}
T=  \int_c \left( \frac{1}{2} A_{\rho} \left| \dot{\boldsymbol  \varphi} \right| ^2 + \frac{1}{2} \sum_{i=1}^{2} M^i_{\rho}\left| \dot{\mathbf{d}}_i \right| ^2 \right) ds, \label{T}
\end{align}
where $A_{\rho}$ is the mass density per reference arc-length and $M^i_{\rho}$ are the principle mass moments of inertia of cross section. The internal potential energy is computed by an integration of the beam {free energy} density $ \Omega^b$ in Eq.~(\ref{Ob}) over the beam center line
\begin{align}
V(\mathbf{q}) = \int_c \Omega^b(\boldsymbol{\Gamma}, \mathbf{K}, \boldsymbol{\Xi}, \boldsymbol{\Theta}) ds.
\end{align}

According to the {kinematic} assumptions in geometrically exact beams, the directors have to fulfill the {orthonormality} constraints, {which serve as the internal constraints $\mathbf{g}^{\rm int}$. The Dirichlet boundary conditions and the mechanical joints are treated as the external constraints $\mathbf{g}^{\rm ext}$. The total constraints reads}
\begin{align}{
	\mathbf{g}( \mathbf{q})=\begin{bmatrix}\mathbf{g}^{\rm int}( \mathbf{q}) & \mathbf{g}^{\rm ext}( \mathbf{q})
	\end{bmatrix}^T=\mathbf{0}.}
\end{align}
The contact constraints are treated by means of the gap function $\mathbf{g}^c(\mathbf{q})$ and the Lagrange multiplier $\boldsymbol{\lambda}^c$ in Eq.~(\ref{dAp2}), which is complemented with the Kuhn–Tucker condition, see \citep{wriggers2006computational} for instance,
\begin{align}
	\mathbf{g}^c(\mathbf{q}) \geqslant 0, \qquad \boldsymbol{\lambda}^c \leqslant 0, \qquad 	\mathbf{g}^c(\mathbf{q}) \cdot \boldsymbol{\lambda}^c =0.
\end{align}
{In this work, $\mathbf{g}^c(\mathbf{q})$ contains the normal gap function only whereas the tangential effects like friction are neglected. Nevertheless, viscosity is considered} in the external source term $\mathbf{f}^{\rm ext}$ in Eq.~(\ref{dAp2}) containing the viscoelastic forces and electric charges
\begin{align}
	\mathbf{f}^{\rm ext}= \begin{bmatrix} \mathbf{f}^{\rm v}(\mathbf{q},\dot{\mathbf{q}}) & \mathbf{Q} \end{bmatrix}^T,
\end{align}
where $\mathbf{Q}$ is a vector containing the electric charges as source term for the electrical degrees of freedom $\boldsymbol{\phi}$. The viscoelastic forces in the dielectric material is accounted for by using the Kelvin-Voigt model, see \citep{linn2013geometrically},
\begin{align}
	\mathbf{f}^{\rm v}=\int_c \left( \mathbf{M}^{\rm v} \frac{\partial \boldsymbol{\Gamma}}{\partial \mathbf{q}} + \mathbf{N}^{\rm v} \frac{\partial \mathbf{K}}{\partial \mathbf{q}} \right)  ds, \qquad
		\mathbf{M}^{\rm v} = \mathbf{D}_5 \dot{\boldsymbol{\Gamma}}, \quad \mathbf{N}^{\rm v}=\mathbf{D}_6 \dot{\mathbf{K}}, \label{damp}
\end{align}
where $\mathbf{D}_5$ and $\mathbf{D}_6$ are damping parameters.

\subsection{Constrained discrete Euler–Lagrange equations with null space projection}
In this work, the electromechanically coupled beam dynamics is {approximated} within the constrained discrete variational scheme with the null space projection \citep{leyendecker2008variational}. The beam is first spatially discretized with the 1D finite elements \citep{leyendecker2004energy}, where one-dimensional Lagrange-type linear shape functions $N_I$ are applied to interpolate beam configuration $\mathbf{q}_e$ in the element
\begin{align}
\mathbf{q}_e = \sum_{I=1}^{2} N_I(\xi)\mathbf{q}_I.
\end{align}
In this case, the beam directors are directly discretized in space together with the beam centroids and the electrical configurations, see \citep{romero2002objective} for instance. Then the variational integration scheme, see e.g. \citep{leyendecker2008variational}, is applied to temporally discretize the action of the dynamical system, by which the {good} long term energy {behavior} can be obtained. In the variational integration scheme, the action integral of Lagrangian within the time interval $(t_n,t_{n+1})$ is approximated with the discrete Lagrangian $L_d$ as
\begin{align}
\int_{t_n}^{t_{n+1}} L(\mathbf{q},  \dot{\mathbf{q}})dt \approx L_d(\mathbf{q}_n,\mathbf{q}_{n+1}) = \Delta t L(\frac{\mathbf{q}_{n+1}+\mathbf{q}_n}{2},\frac{\mathbf{q}_{n+1}-\mathbf{q}_n}{\Delta t}),
\end{align}
where the discrete Lagrangian $L_d$ is computed by applying the finite difference approximation to the velocity $\dot{\mathbf{q}}$ and the midpoint {rule} to the configuration $\mathbf{q}$. The action integral of contact part is approximated with the mid-point rule
\begin{align}
	\int_{t_n}^{t_{n+1}}  \mathbf{g}^{cT}(\mathbf{q})\cdot \boldsymbol\lambda^c dt \approx \frac{\mathbf{g}^c(\mathbf{q}_n) \cdot \boldsymbol{\lambda}_n^c + \mathbf{g}^c(\mathbf{q}_{n+1}) \cdot \boldsymbol{\lambda}_{n+1}^c}{2},
\end{align}
where the time step $\Delta t$ does not appear explicitly but is included in the multipliers instead since the contact constraints may not hold in the full time interval. The midpoint rule is applied to the action integral of holonomic constraints as well. After the temporal discretization, the discrete Euler–Lagrange equations can be obtained by taking the variation of the discrete action and requiring stationarity. To eliminate the internal constraint forces from the system, see e.g. \citep{betsch2006discrete}, the internal discrete null space matrices $\mathbf{P}_d^{\rm int}(\mathbf{q}_n)$ are applied to the discrete Euler–Lagrange equations leading to
\begin{align}
	\mathbf{P}_d^{{\rm int},T}(\mathbf{q}_n) \left[ \frac{\partial L_d(\mathbf{q}_{n-1}, \mathbf{q}_{n})}{\partial \mathbf{q}_{n}} + \frac{\partial L_d\left( \mathbf{q}_{n}, \mathbf{q}_{n+1}\right) }{\partial \mathbf{q}_{n}} + \mathbf{G}^{{\rm ext},T}(\mathbf{q}_n) \boldsymbol{\lambda}^{\rm ext}_n+ \mathbf{G}^{cT}(\mathbf{q}_n) \boldsymbol{\lambda}^c_n+ \mathbf{f}_n^{\rm ext-} + \mathbf{f}_{n-1}^{\rm ext+} \right] &= \mathbf{0}, \nonumber\\
\mathbf{g}(\mathbf{q}_n) & = \mathbf{0}, \nonumber\\
\mathbf{g}^c(\mathbf{q}_n) \geqslant \mathbf{0}, \qquad \boldsymbol{\lambda}^c_n \leqslant \mathbf{0}, \qquad \mathbf{g}^c(\mathbf{q}_n) \cdot \boldsymbol{\lambda}^c_n & = \mathbf{0}, \label{fwd}
\end{align}
where $\mathbf{G}^{\rm ext}(\mathbf{q}_n)=\frac{\partial \mathbf{g}_n^{\rm ext}(\mathbf{q}_n)}{\partial \mathbf{q}_n}, \mathbf{G}^c(\mathbf{q}_n)=\frac{\partial \mathbf{g}_n^c(\mathbf{q}_n)}{\partial \mathbf{q}_n}$ and the internal null space matrix is given by
\begin{align}
	\mathbf{P}_d^{\rm int}(\mathbf{q}_n)=
	\begin{bmatrix}                           
		\mathbf{I} & \mathbf{0} & \mathbf{0}\\                                               
		\mathbf{0} & -\hat{\mathbf{d}}_{1,n} & \mathbf{0}\\   
		\mathbf{0} & -\hat{\mathbf{d}}_{2,n} & \mathbf{0}\\  
		\mathbf{0} & -\hat{\mathbf{d}}_{3,n} & \mathbf{0}\\
		\mathbf{0}& \mathbf{0}& \mathbf{I}
	\end{bmatrix}
\end{align}
with $\hat{\mathbf{d}}_{i,n}$ being the skew-symmetric matrix corresponding to the director vector $\mathbf{d}_{i,n}$ at $t_n$ and $\mathbf{I}$ being the 3 by 3 identity matrix. {Note that in Eq.~(\ref{fwd}), the Lagrange multipliers $\boldsymbol{\lambda}^{\rm ext}_n$ for the external constraints are not eliminated from the system by null space projection. Given the initial conditions, the approximated trajectories of the dynamical system can be obtained by solving the discretized forward dynamics problem described in Eq.~(\ref{fwd}).}

\section{Optimal control formulation of the coupled multibody system}
The optimal control of the electromechanically coupled multibody dynamics system is {represented by} the constrained optimization problem, where the control objective is minimized {subject to} the equality constraints and the inequality constraints, {see \citep{leyendecker2010discrete}}. Given an initial and {a final state} of the coupled multibody dynamics system, the optimal trajectory can be computed by solving the constrained optimization problem. In case of the dielectric elastomer actuated multibody system, the trajectories of both mechanical and electrical configurations, the Lagrange multipliers and the electric charges are treated as optimization variables
\begin{align}
	\mathbf{x}=
	\begin{bmatrix} \mathbf{q}_0 &  ... &  \mathbf{q}_{N} &  \boldsymbol{\lambda}_0^{\rm ext} & ... \boldsymbol{\lambda}_{N-1}^{\rm ext} & \boldsymbol{\lambda}_0^c &  ... \boldsymbol{\lambda}_{N-1}^c & \mathbf{Q}_0 & ... & \mathbf{Q}_{N-1} \end{bmatrix}, \label{x}
\end{align}
where $N$ is the number of time steps. The control objective is to {ensure} minimum variation of the electric potential over time {and can be} written as
\begin{align}
	J(\mathbf{x})=\sum_{n=1}^{N} \sum_{I=1}^{3 \times N_{\rm node}}  \left( \phi^I_n-\phi^I_{n-1}\right)^2 \label{obj}
\end{align}
with $N_{node}$ the number of beam nodes and $\phi^I$ the $I$-th electrical degree of freedom of the dynamical system. The discrete Euler-Lagrange equations, the internal constraints, the initial and final conditions together with the equality part of contact constraints are treated as equality constraints
\begin{align}
	n=0: \qquad & \mathbf{q}_0^{\rm mech} = \bar{\mathbf{q}}_0^{\rm mech} \label{ini}\\
	&\bar{\mathbf{p}}_0 - \mathbf{p}_0^-(\mathbf{q}_0, \mathbf{q}_1, \boldsymbol{\lambda}_0, \boldsymbol{\lambda}^c_0, \mathbf{Q}_0) =\mathbf{0}\\
	&\mathbf{g}(\mathbf{q}_1) =\mathbf{0}  \\
	&\mathbf{g}^c(\mathbf{q}_1) \odot \boldsymbol{\lambda}_0^c=\mathbf{0}\\
	& \nonumber\\
	n=1...N-1:\qquad &\mathbf{p}_n^+(\mathbf{q}_{n-1}, \mathbf{q}_n, \boldsymbol{\lambda}_n, \boldsymbol{\lambda}_n^c, \mathbf{Q}_n)  - \mathbf{p}_n^-(\mathbf{q}_n, \mathbf{q}_{n+1}, \boldsymbol{\lambda}_n, \boldsymbol{\lambda}^c_n, \mathbf{Q}_n) =\mathbf{0}  \\
	&\mathbf{g}(\mathbf{q}_{n+1}) = \mathbf{0}  \\
	&\mathbf{g}^c(\mathbf{q}_{n+1}) \odot \boldsymbol{\lambda}_n^c= \mathbf{0}  \\
	&\nonumber \\
	n=N: \qquad & \mathbf{p}_N^+(\mathbf{q}_{N-1}, \mathbf{q}_N, \boldsymbol{\lambda}_N, \boldsymbol{\lambda}_N^c, \mathbf{Q}_N)  - \bar{\mathbf{p}}_N=\mathbf{0}\\
	\qquad & \mathbf{q}_N^{\rm mech}  = \bar{\mathbf{q}}_N^{\rm mech}, \label{final}
\end{align}
{ where the discrete conjugate momenta $(\mathbf{p}_n^-,\mathbf{p}_n^+)$ are obtained from the discrete Legendre transformation}
\begin{align}
& \mathbf{p}_n^-(\mathbf{q}_n, \mathbf{q}_{n+1}, \boldsymbol{\lambda}_n)=-D_1L_d(\mathbf{q}_n,\mathbf{q}_{n+1}) +\frac{1}{2}\mathbf{G}^T_c(\mathbf{q}_n)\boldsymbol{\lambda}_n^c + \frac{1}{2}\mathbf{G}^{{\rm ext}, T}(\mathbf{q}_n)\boldsymbol{\lambda}_n^{\rm ext} -\mathbf{f}_n^{\rm ext -},\\
	& \mathbf{p}_n^+(\mathbf{q}_{n-1}, \mathbf{q}_{n}, \boldsymbol{\lambda}_n)=D_2L_d(\mathbf{q}_{n-1},\mathbf{q}_{n}) -\frac{1}{2}\mathbf{G}^T_c(\mathbf{q}_n)\boldsymbol{\lambda}_n^c -\frac{1}{2}\mathbf{G}^{{\rm ext}, T}(\mathbf{q}_n)\boldsymbol{\lambda}_n^{\rm ext} +\mathbf{f}_{n-1}^{\rm ext+}
\end{align}
{with $\bar{\mathbf{q}}_0^{\rm mech} $ and $\bar{\mathbf{q}}_N^{\rm mech} $ the prescribed configurations, $\bar{\mathbf{p}}_0$ and $\bar{\mathbf{p}}_N$ the given initial and final momenta and $\odot$ denoting the element-wise multiplication. Note that only the mechanical configurations are involved in the initial and the final condition whereas the electrical configurations are treated as unknown variables. From Eq.~(\ref{fwd}), the inequality constraints
\begin{align}
	n=1...N-1: \qquad  &\mathbf{g}_n^{c}( \mathbf{q}_n) \geqslant \mathbf{0},\\
	& \boldsymbol\lambda_n^{c}  \leqslant \mathbf{0}
\end{align}
are amended to the problem. To solve the constrained optimization problem above, the IPOPT solver and the automatic differentiation tool CasADi \citep{Andersson2019} are used.}

\section{Numerical examples}
To demonstrate the effectiveness of the {free energy function} as well as the optimal control formulations for multibody dynamics presented above, three numerical examples are investigated in this section. Throughout all the examples, the DEA-beams are controlled by the electric charges. The material parameters of the dielectric elastomer used in this work are shown in Table 1. The damping parameters in Eq.~(\ref{damp}) are given as $\mathbf{D}_5 =\mathbf{D}_6= \eta\mathbf{I}_{3\times3}$ with parameter $\eta$. All of the beams are straight with square cross sections in the reference configuration.

\begin{table}[htb!]
	\centering
	\caption{Material parameters} 
	\begin{tabular}{cccccccc}
		\hline
		$\rho$& $E$ &$G$ & $c_1$ & $c_2$  & $\eta$\\
		\hline
		$g/mm^3$ & $Mpa$ & $Mpa$ & $N/V^2$ & $N/V^2$ &-\\
		\hline
		$0.1$ &  $654.9$  & $233$ &  $5\times 10^{-8}$  &  $1\times 10^{-3}$ & 500\\
		\hline
	\end{tabular}
\end{table}

\subsection{Cantilever beam}
In this example, we consider the optimal control of the bending of an electromechanically coupled cantilever beam. The geometry of the beam in the reference configuration is given as: length $l=10mm$ and width $b=2mm$. The beam is composed of 10 stacked DEA cells, which are discretized with two-node finite elements. The electric charges are applied on the beam nodes, where the electrodes are placed and the thickness of the electrode connecting two adjacent cells is neglected. The configuration of the straight beam is specified as the initial condition in Eq.~(\ref{ini}), whereas a bending state of the beam is specified as the final mechanical condition in Eq.~(\ref{final}). The control objective is to minimize the variation of the electric potential over time as shown in Eq.~(\ref{obj}).

Before solving the optimization problem, the optimization variables are initialized such that the optimization process is not too expensive and the initial guess is not too far from the expected solution. The mechanical variables in $ (\mathbf{q}_1, ... , \mathbf{q}_{N-1})$ is initialized with the linear increment from the initial mechanical configuration $\mathbf{q}_0^{\rm mech}$ to the manually defined final one $\mathbf{q}_N^{\rm mech}$. The electrical variables in $ (\mathbf{q}_1, ... , \mathbf{q}_{N-1})$ are initialized with constants $\boldsymbol{\phi}_I=\begin{bmatrix} 1 & 0 &0 \end{bmatrix}$ for nodes $I=1,3,5,...$ and  $\boldsymbol{\phi}_I=\begin{bmatrix} 1 & 0 &-1 \end{bmatrix}$ for nodes $I=2,4,6,...$. The Lagrange multipliers $(\boldsymbol{\lambda}_0^{\rm ext}, ... \boldsymbol{\lambda}_{N-1}^{\rm ext}, \boldsymbol{\lambda}_0^c, ... \boldsymbol{\lambda}_{N-1}^c )$ and  the electric charges $ (\mathbf{Q}_0, ... , \mathbf{Q}_{N-1})$ are initialized with zeros.

\begin{figure}[htb!]
	\centering
	\subfigure[$t=0ms$]{\label{.}\includegraphics[width=.4\textwidth]{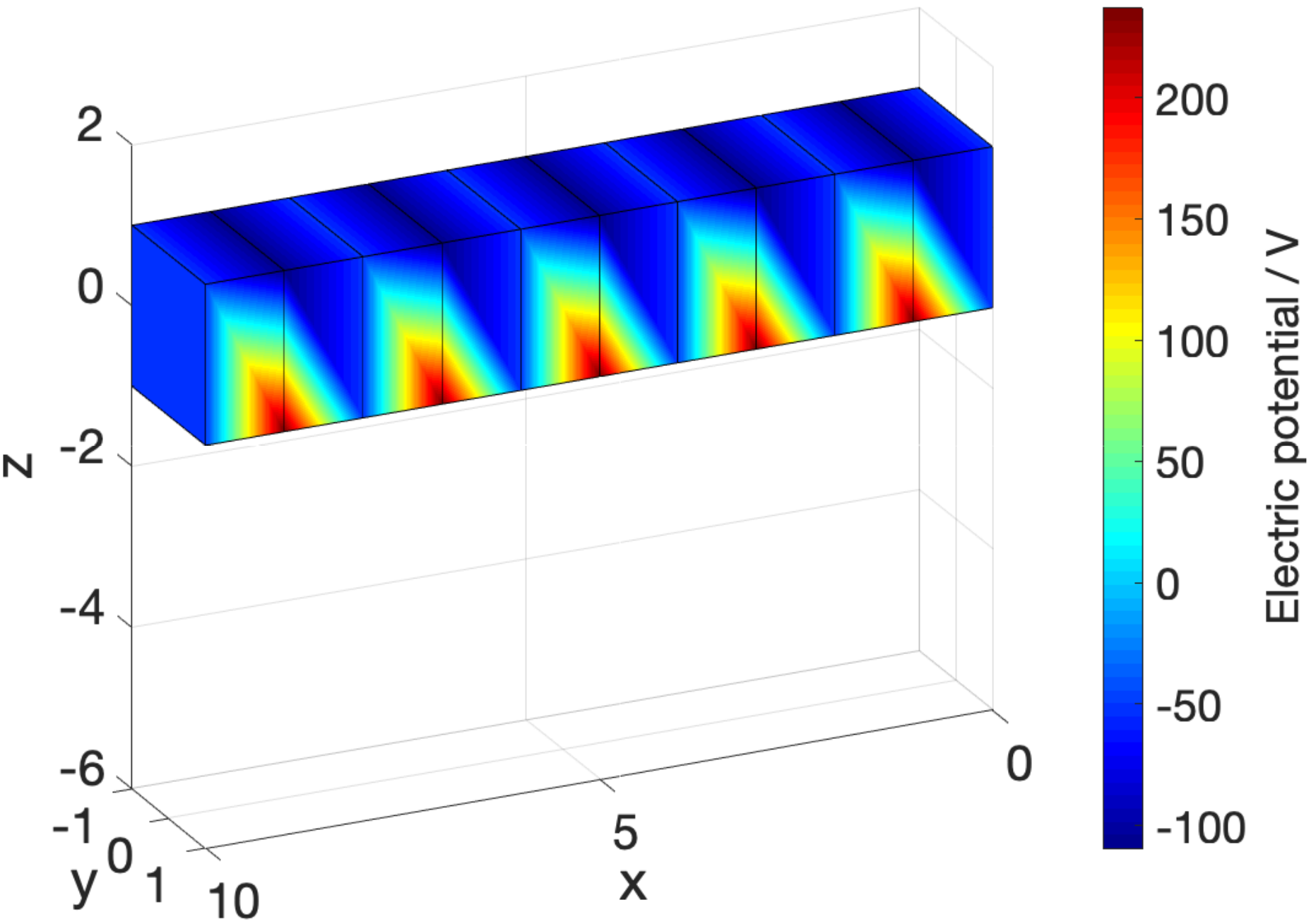}}
	\qquad
	\subfigure[$t=0.2ms$]{\label{.}\includegraphics[width=.4\textwidth]{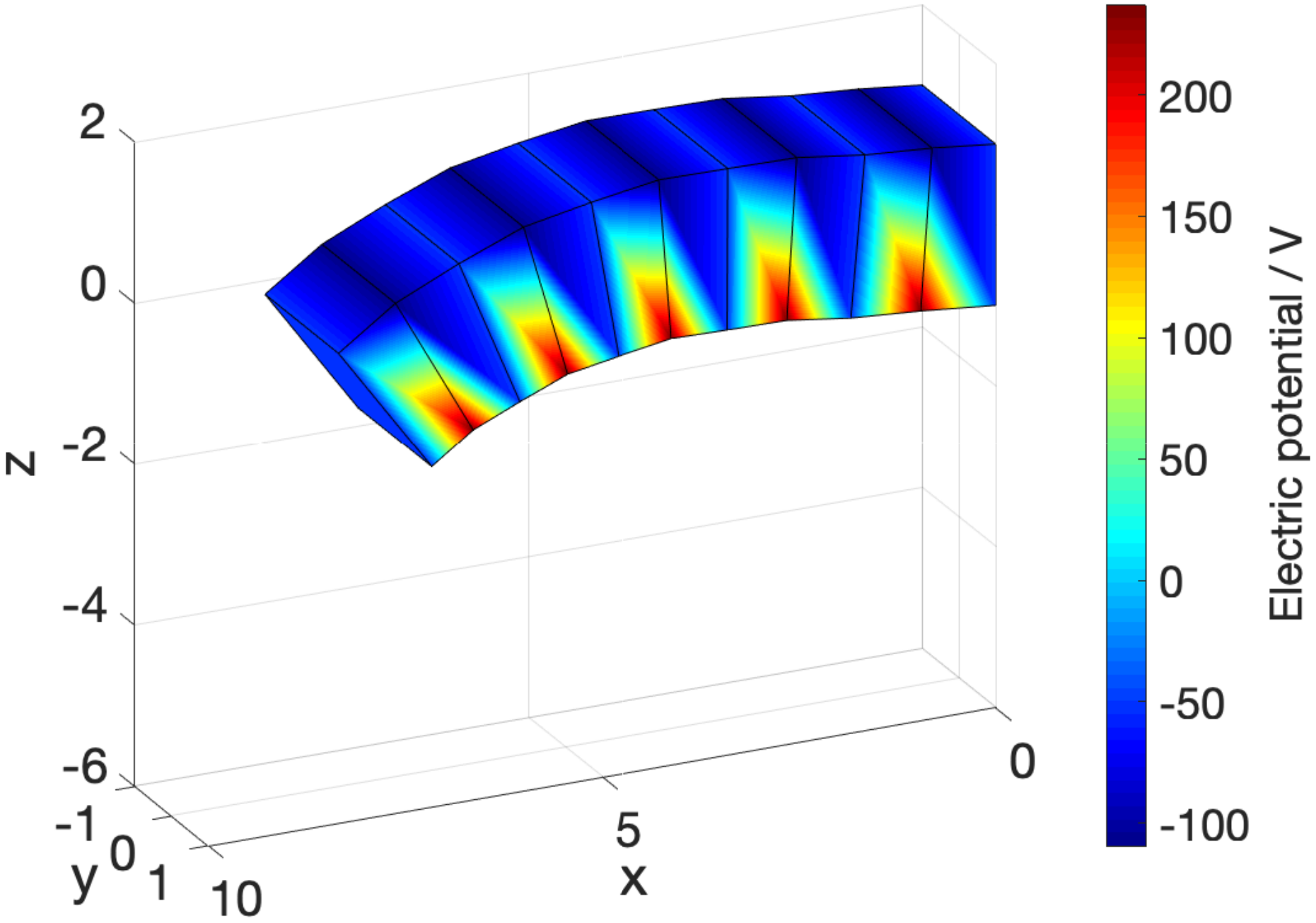}}
	\caption{Configurations of the cantilever beam.}
	\label{b-q}
\end{figure}

\begin{figure}[htb!]
	\centering
	\subfigure[$\varphi_x$ of node $I=11$]{\includegraphics[width=.4\textwidth]{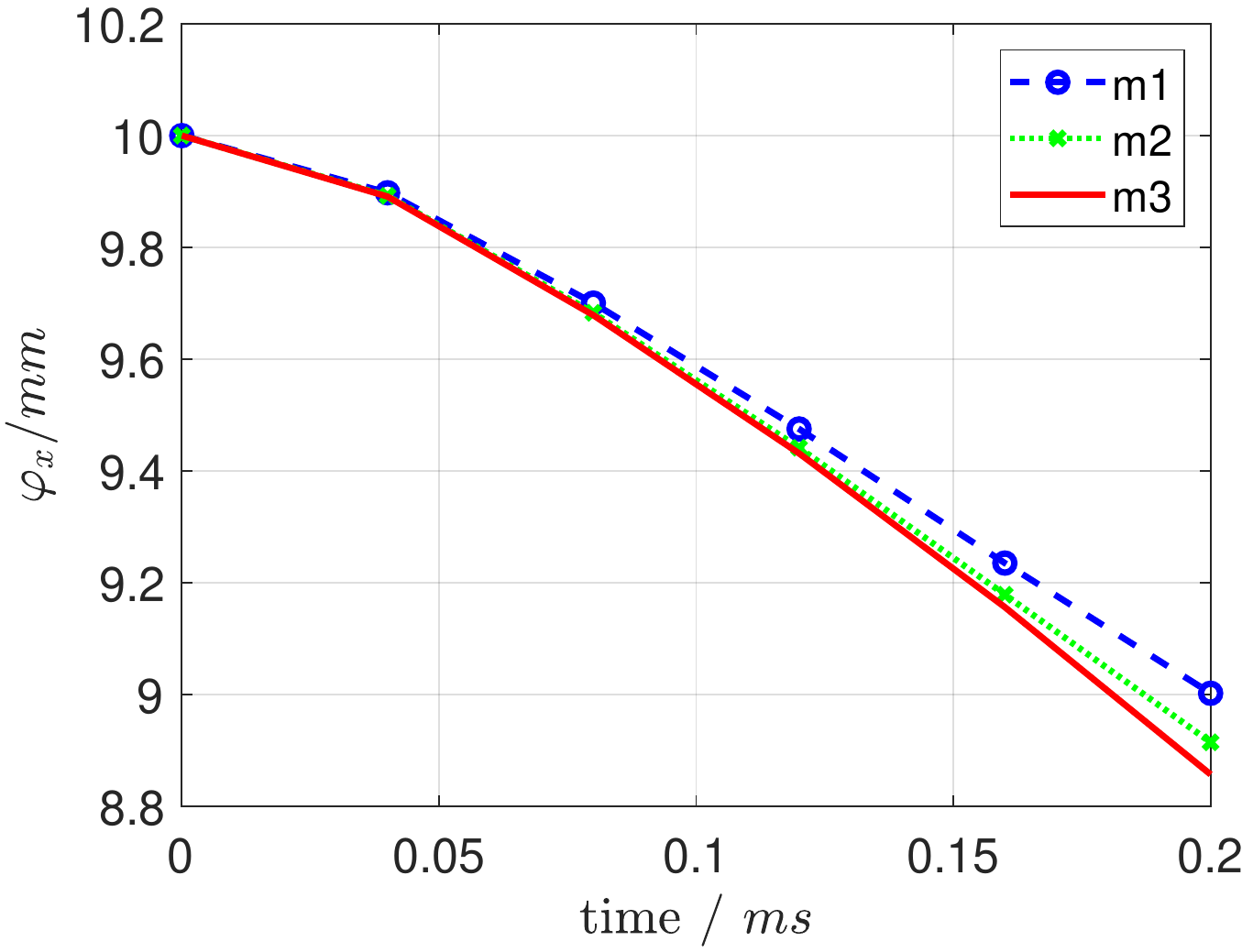}}
	\qquad
	\subfigure[$\varphi_z$ of node $I=11$]{\includegraphics[width=.4\textwidth]{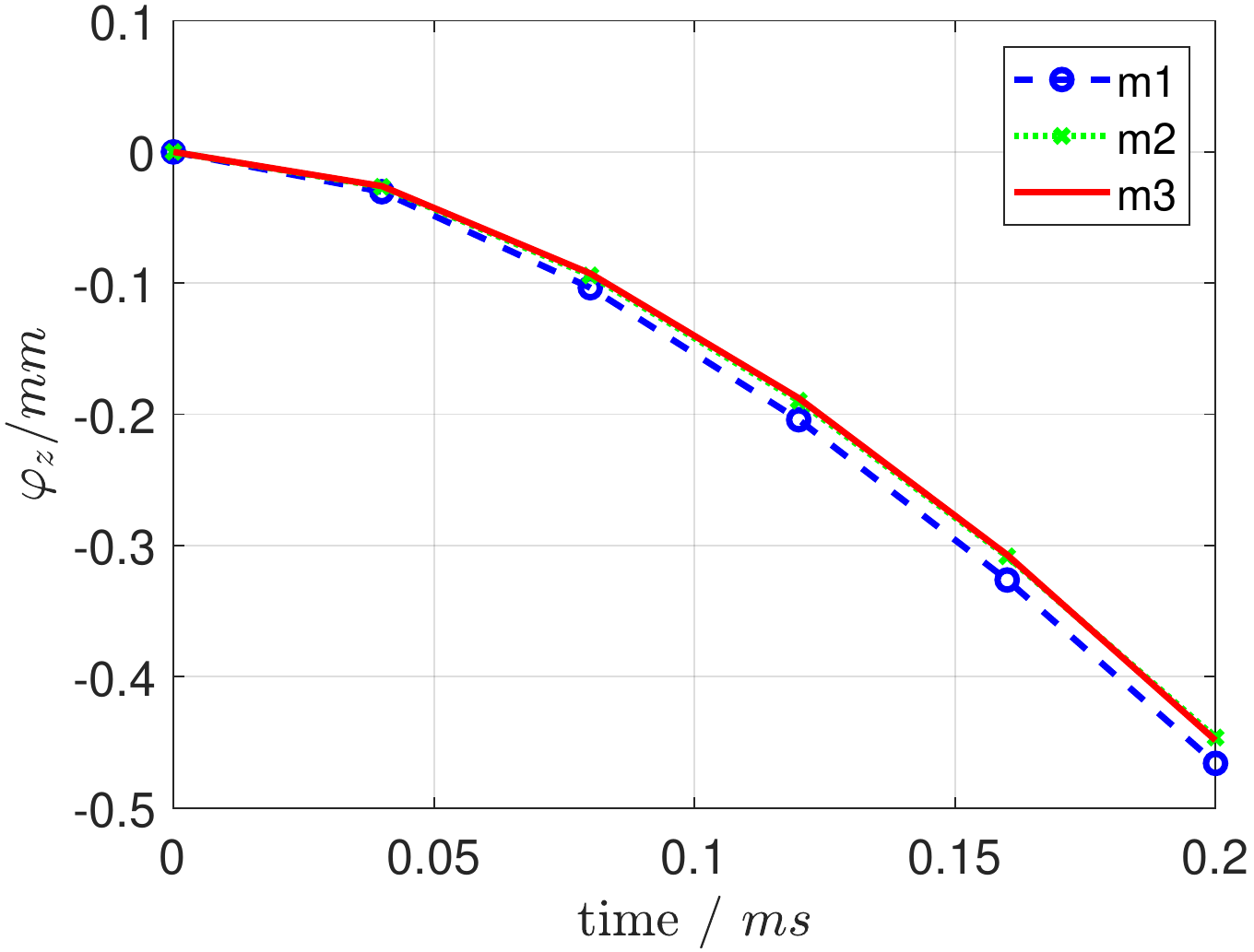}}\\
	\subfigure[$\varphi_x$ of node $I=11$]{\includegraphics[width=.4\textwidth]{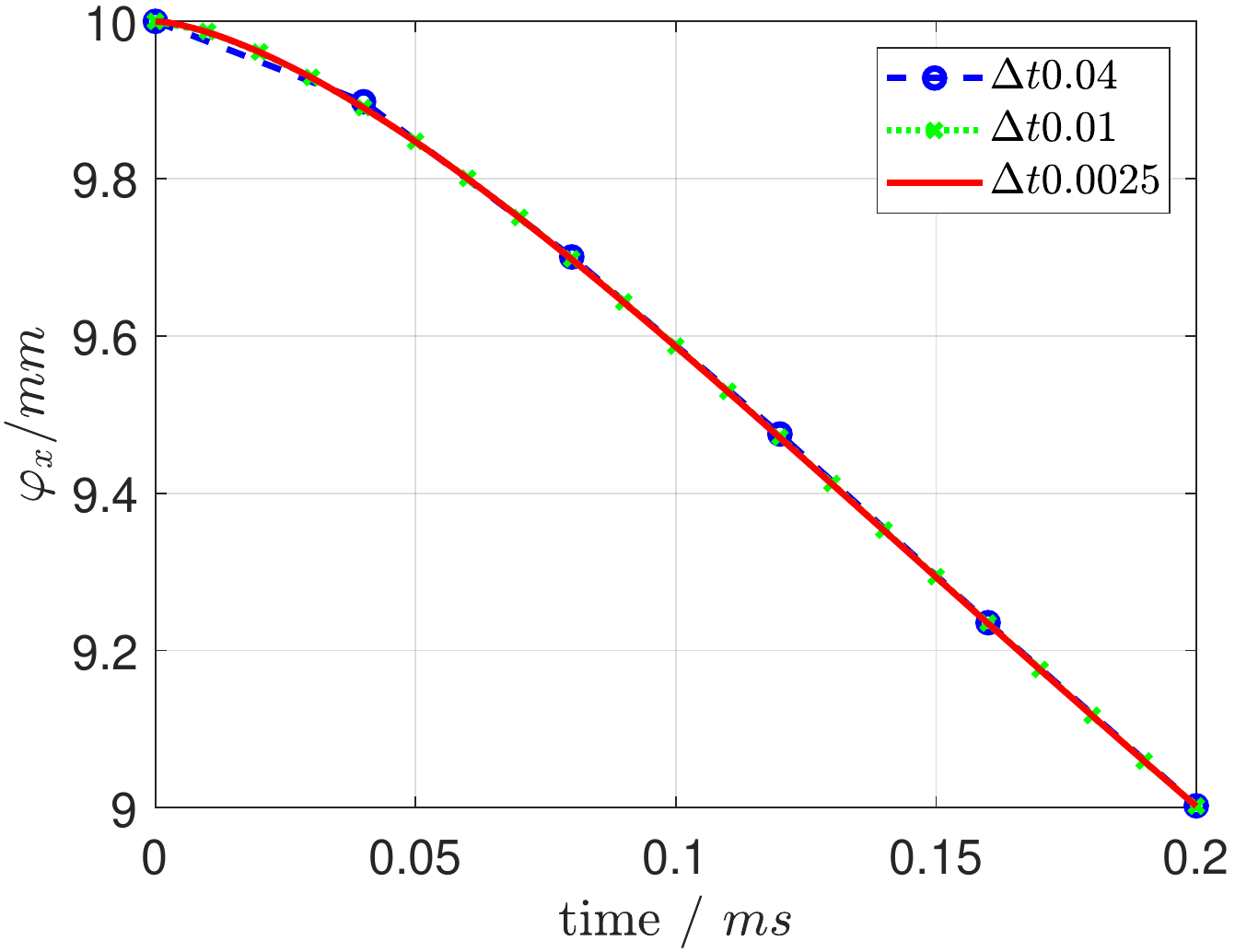}}
	\qquad
	\subfigure[$\varphi_z$ of node $I=11$]{\includegraphics[width=.4\textwidth]{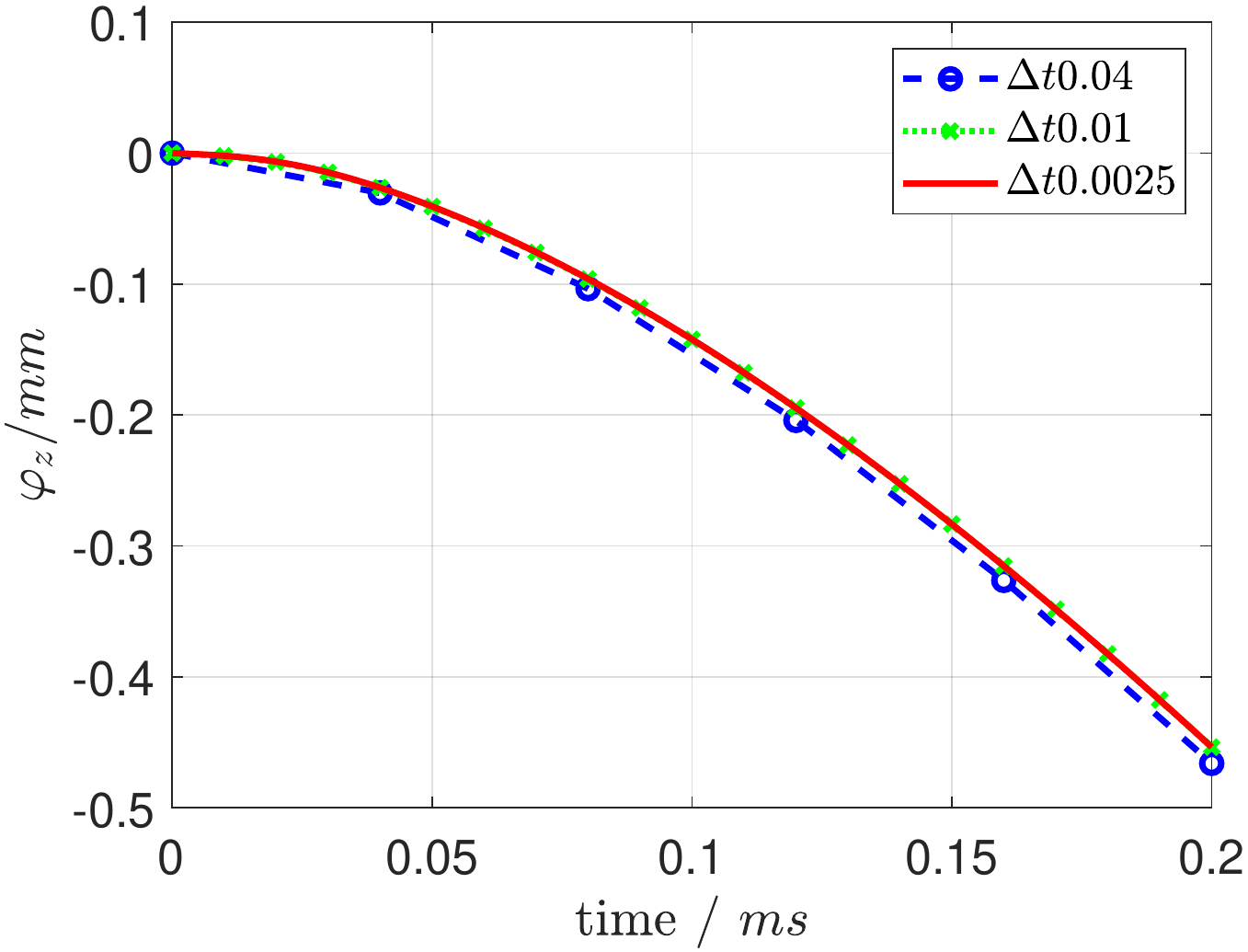}}
	\caption{Motion of the last beam node with different mesh sizes and time steps.}
	\label{b-u}
\end{figure}

\begin{figure}[htb!]
	\centering
	\subfigure[$\phi_o$ and $\alpha$ on the 1st electroid]{\includegraphics[width=.45\textwidth]{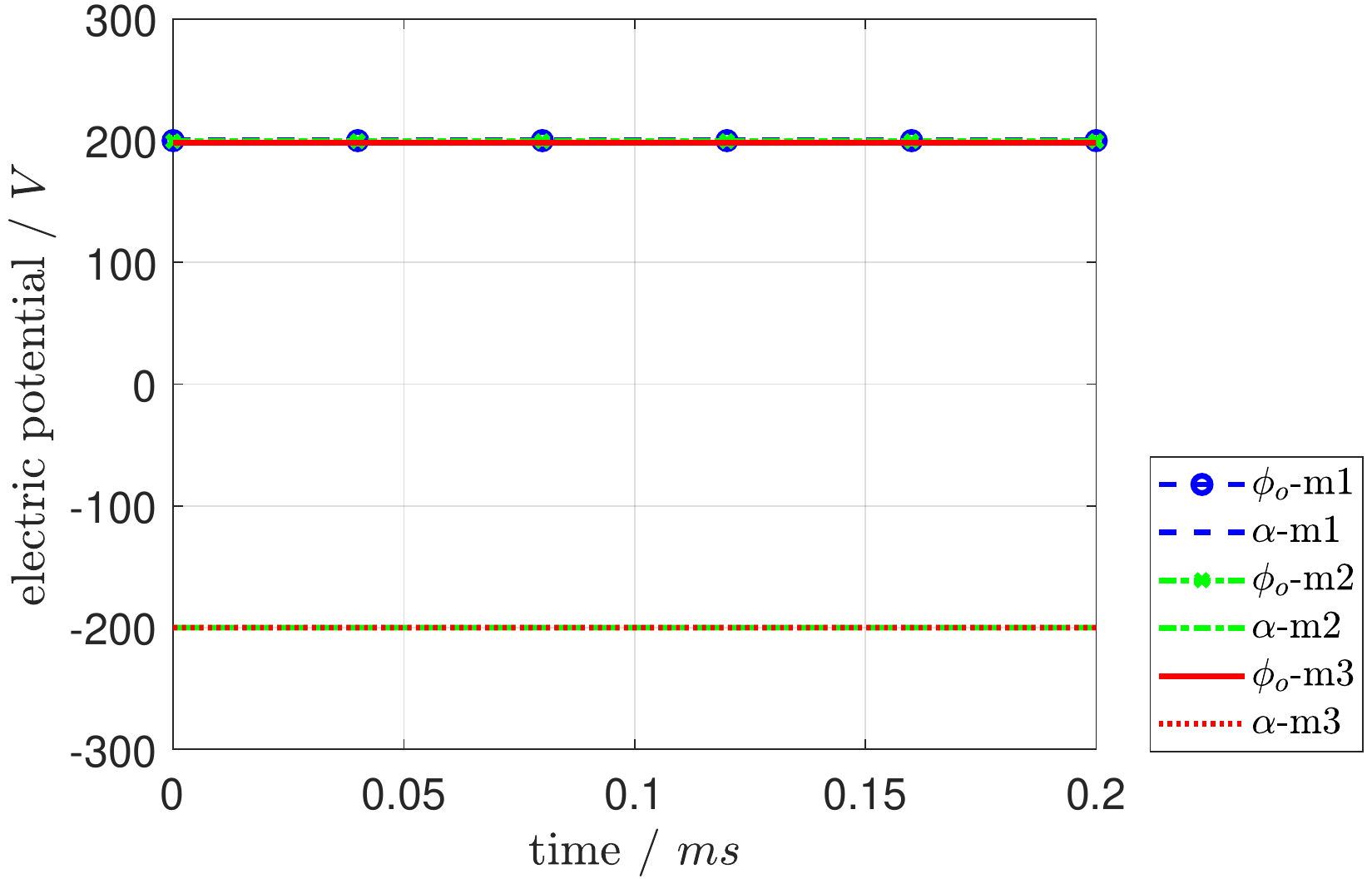}}
	\qquad
	\subfigure[$\phi_o$ and $\alpha$ on the 2nd electroid]{\includegraphics[width=.45\textwidth]{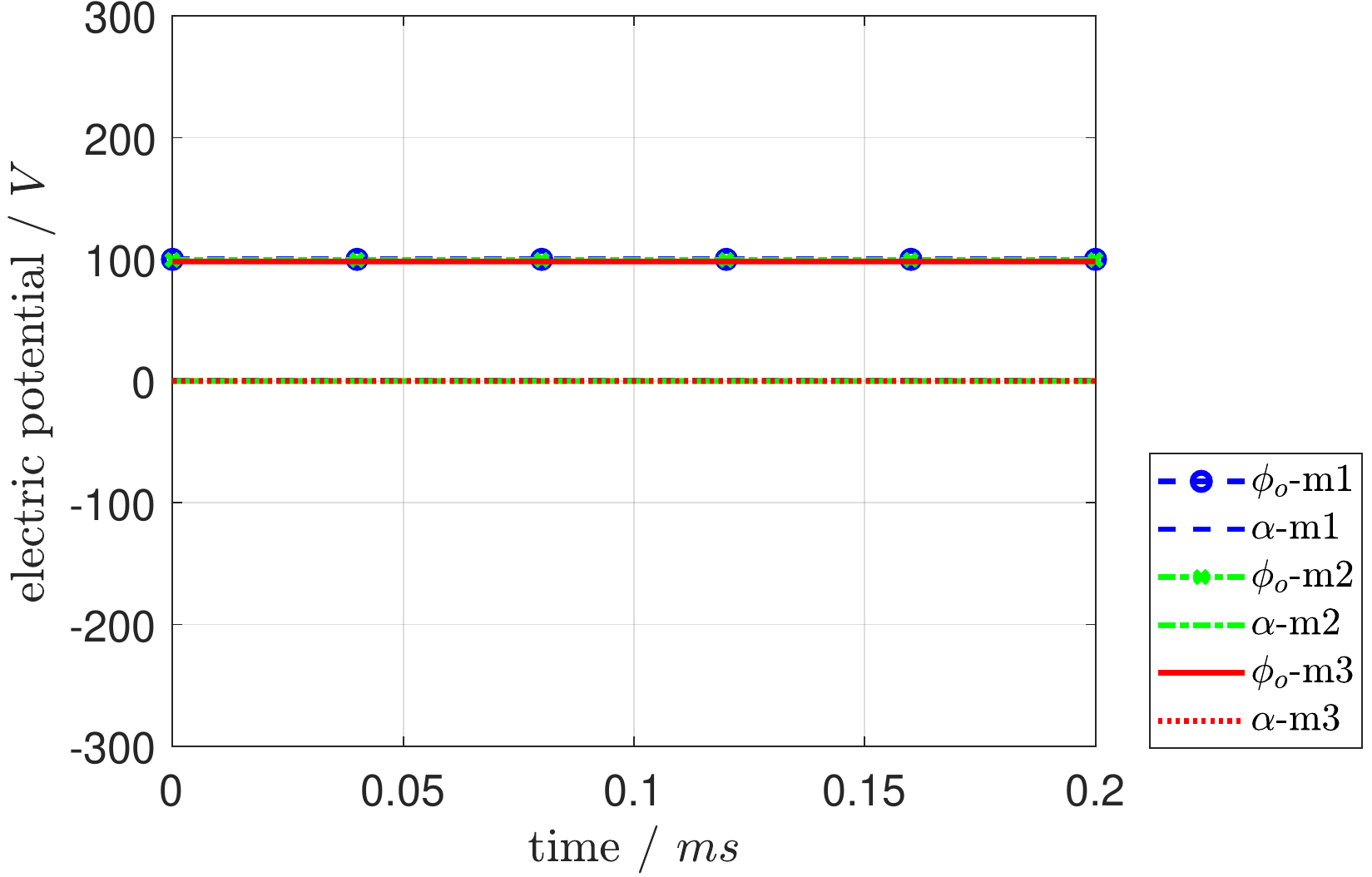}}\\
	\subfigure[$\phi_o$ and $\alpha$ on the 1st electroid]{\includegraphics[width=.45\textwidth]{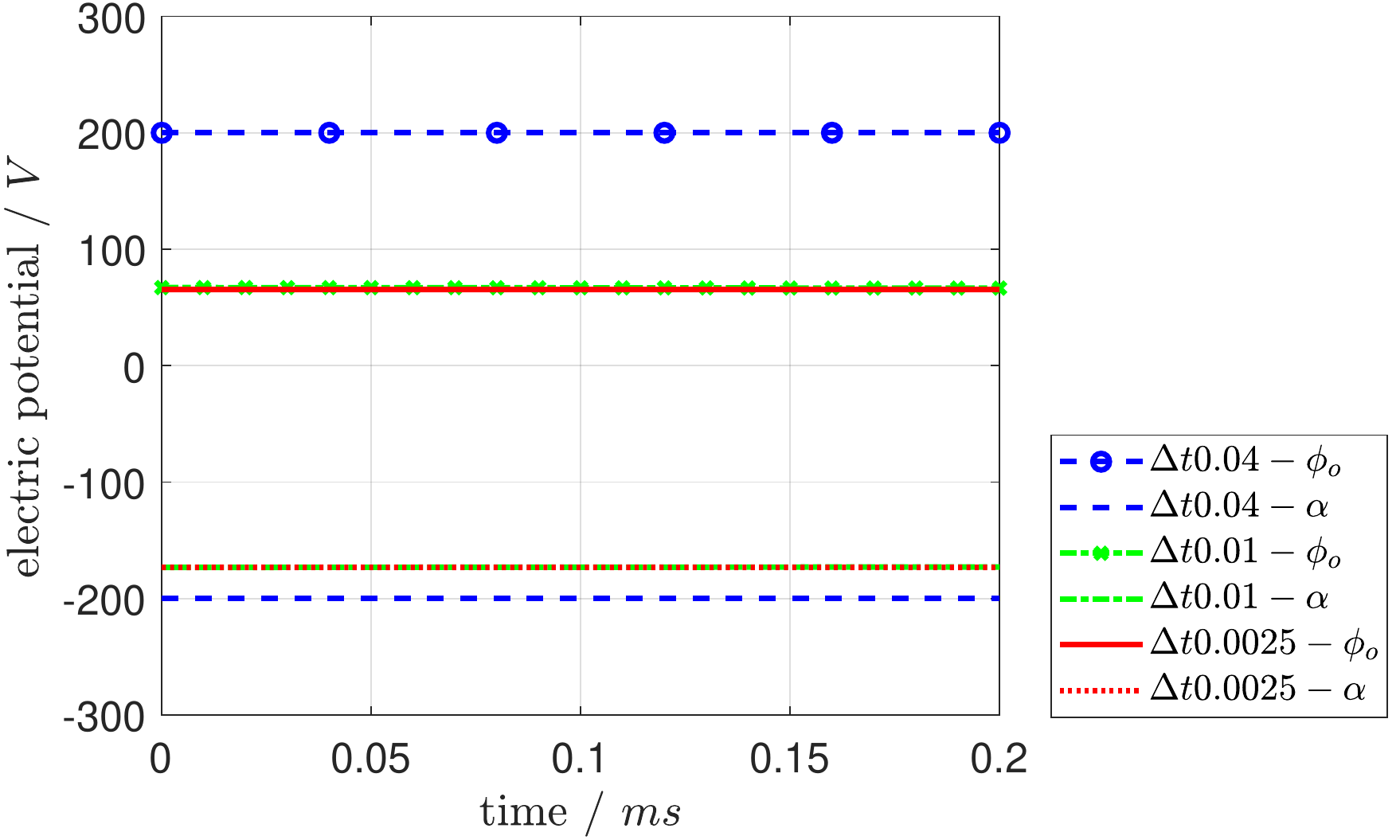}}
	\qquad
	\subfigure[$\phi_o$ and $\alpha$ on the 2nd electroid]{\includegraphics[width=.45\textwidth]{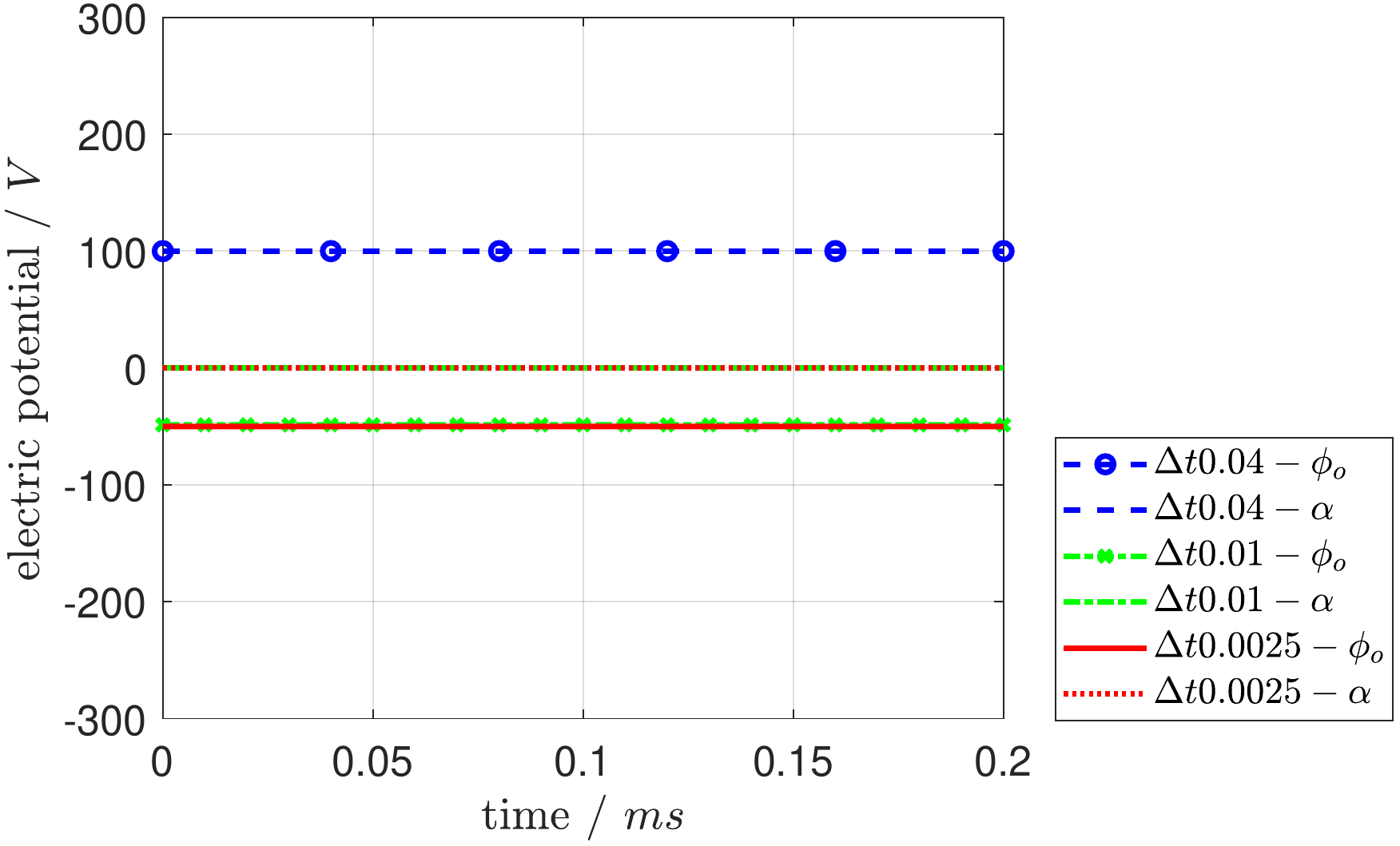}}
	\caption{Electric potential of the last DEA cell on the beam.}
	\label{b-v}
\end{figure}

By setting the  time step as $\Delta t=0.04ms$ and applying one finite element in each DEA cell firstly, the optimal trajectories of the configuration, the Lagrange multipliers and the electric charges of the cantilever beam are obtained as shown in Fig.~\ref{b-q}, where the initial and final configurations of the cantilever beam are plotted. To investigate the influence of mesh size and time step size on the optimal control solution, the optimization problem is solved with different mesh sizes and time steps. The motion of the last beam node for different mesh sizes is compared in Fig.~\ref{b-u}(a) and (b), where one, two and three elements (corresponding to m1, m2 and m3 on the figures) are used in each DEA cell, respectively. The motion of the last beam node for different time steps is shown in Fig.~\ref{b-u}(c) and (d). It can be observed that the deformation of the beam is converged with the decrease of mesh size and time step. The electrical configurations of the last DEA cell on the beam are shown in Fig.~\ref{b-v}. The first electrode of the last DEA cell is located on the last beam node. Another one is on the second to last, the third to last and the fourth to last node for the discretization of m1, m2 and m3, respectively. Fig.~\ref{b-v} shows that the constant electric potential over time is maintained. Additionally, the electric potential converges with the decrease of mesh size and time step.

\subsection{Soft robotic worm}
As an extension of the beam actuator to multibody systems, the optimal control of a soft robotic worm is investigated, where two rigid cubes are connected to an electromechanically coupled beam via revolution joints. The frictional contact between the rigid cubes and the ground is considered by imposing bilateral contact constraints in the normal direction and a friction force in the tangential direction. The frictional contact force is approximated with a Heaviside function, $F_x=\frac{1000}{1+{\rm exp}(20 \dot{\varphi}_x)}-20\dot{\varphi}_x$, such that the worm can march forward after the bending or contraction of the beam is actuated. The locomotion of the soft robotic worm is composed of the actuated phase and a passive stretching phase. In this section, the optimal control problem is solved for the actuated phase only. The stretching happens naturally when the electric boundary conditions are released. In the reference configuration, the beam is straight and has the same size as in the last section. There are 6 DEA cells in the beam, where each DEA cell is discretized with one element. 

The unactuated multibody system with the straight beam is set as the initial condition in Eq.~(\ref{ini}). For the final condition in Eq.~(\ref{final}), only the final position ($x^{rb1}_N$) of the left rigid cube is specified, which gives the marching distance of the actuated phase. The final state of the beam as well as of the right cube are not specified. The time step is set as $\Delta t=0.04ms$ and the total time is set as $t=2ms$. 

\begin{figure}[htb!]
	\centering
	\subfigure[contraction]{\label{.}\includegraphics[width=.33\textwidth]{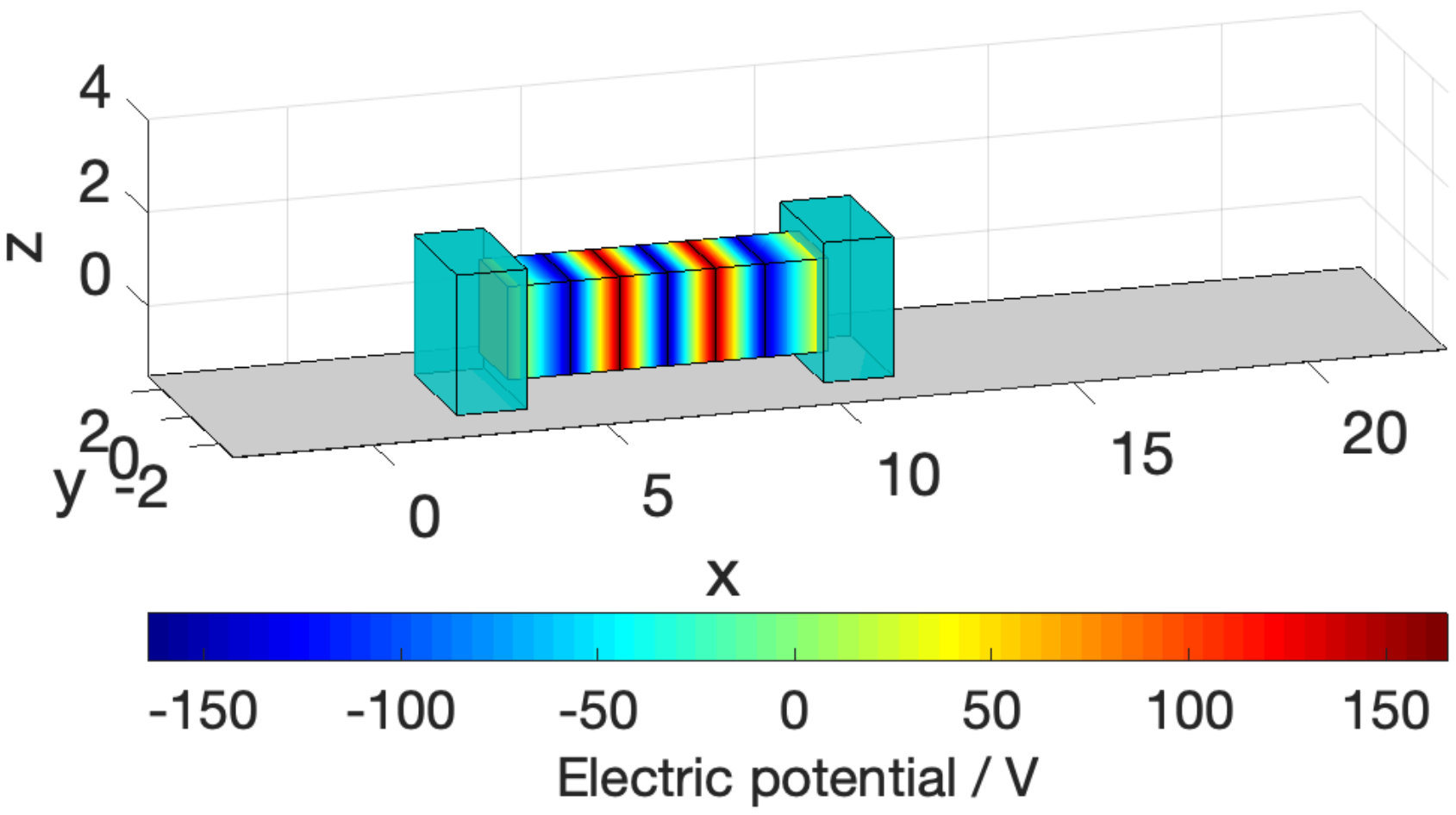}}
	\subfigure[bending]{\label{.}\includegraphics[width=.33\textwidth]{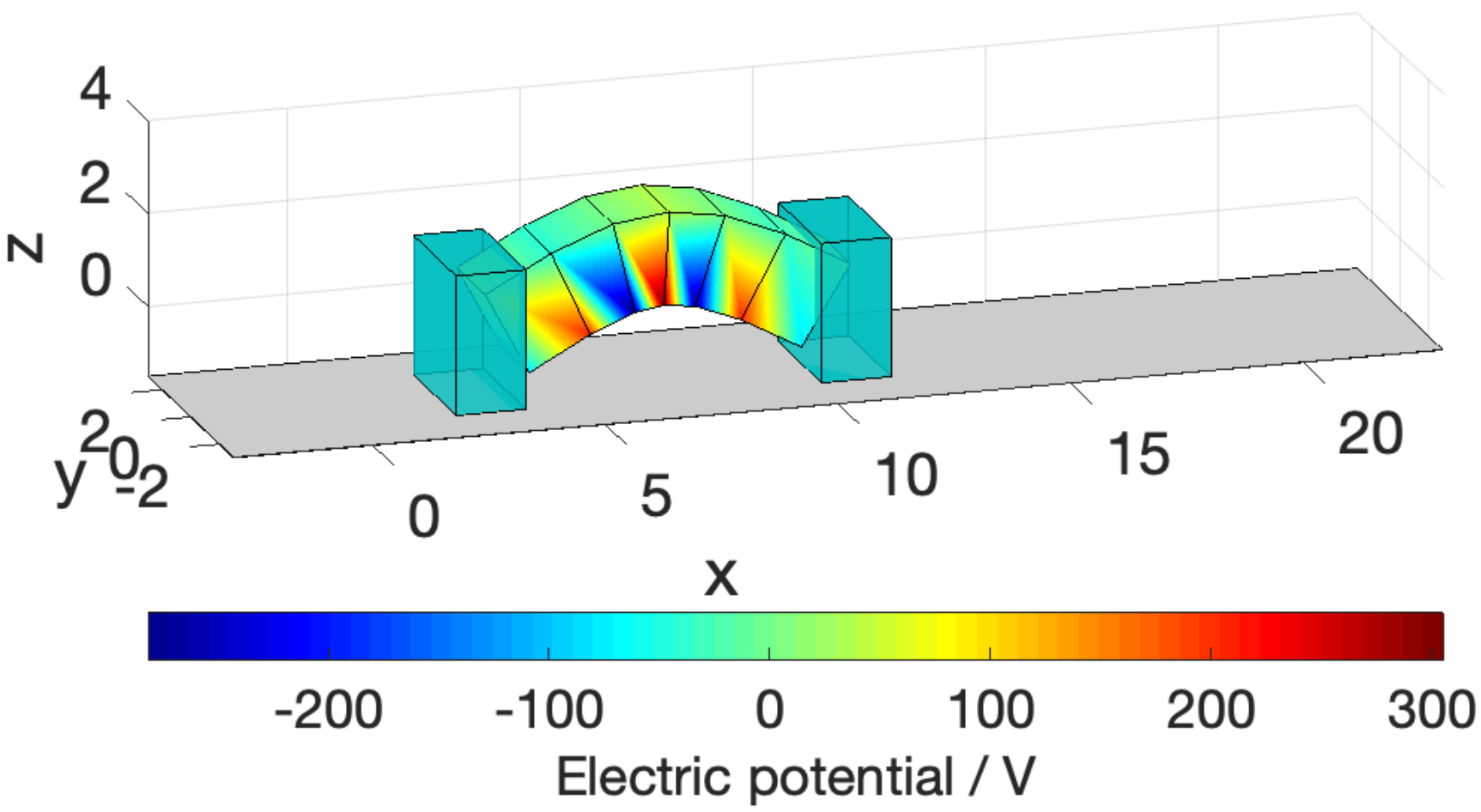}}
	\subfigure[motion of the left rigid cube]{\label{.}\includegraphics[width=.32\textwidth]{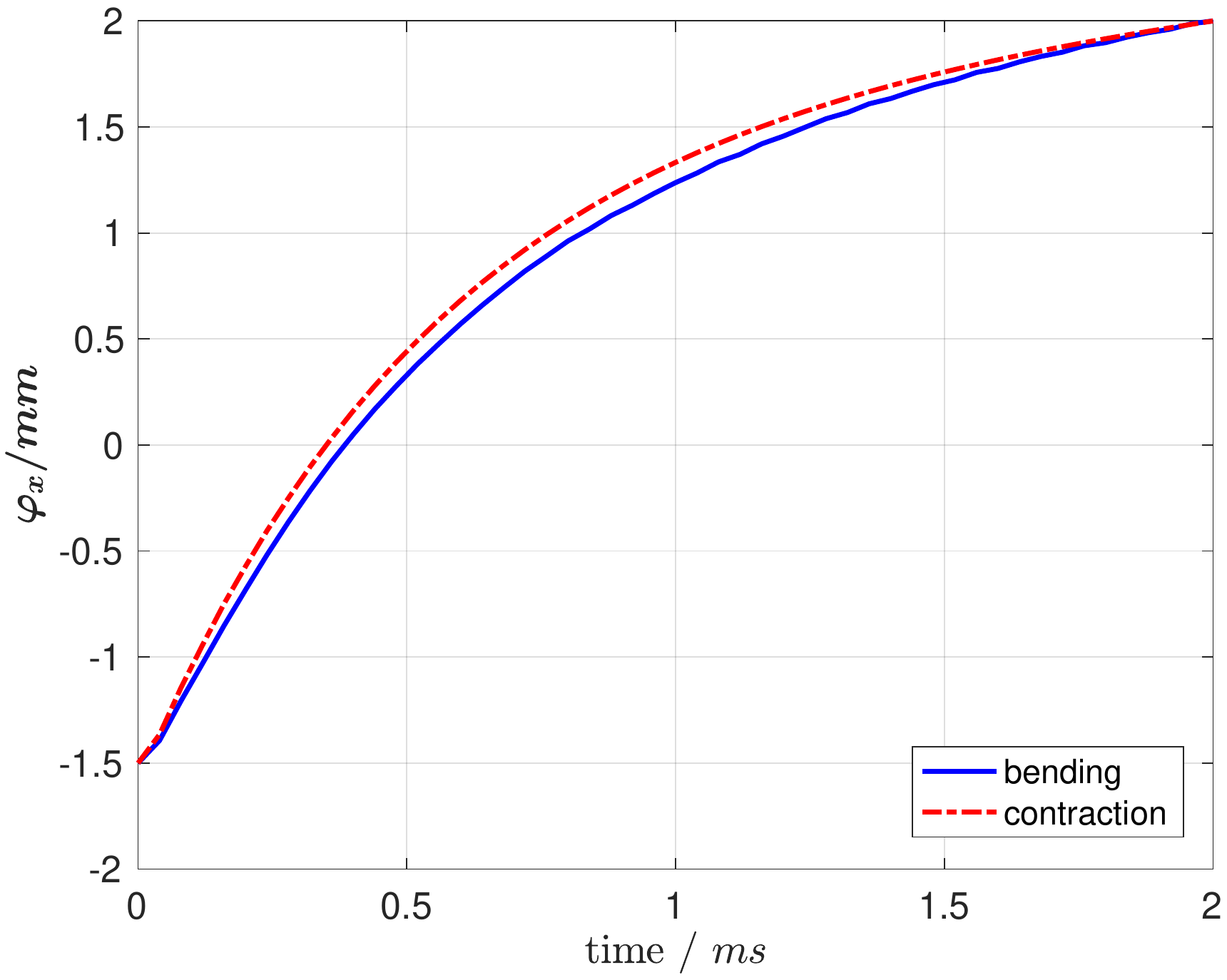}}
	
	\caption{Locomotion of the  soft robotic worm.}
	\label{w-bc}
\end{figure}

The initialization of the optimization variables is crucial for solving the optimal control problem. In the initialization of the optimization variables, the mechanical trajectory is initialized linearly between the initial straight state and a final state. The electrical trajectory is initialized with constant values of 0 and 1. The locomotion of the soft robotic worm can be achieved by actuating either bending or contraction of the beam or combinations thereof. The optimal solution of the trajectories depends on the initialization. When a mechanical configuration of contraction and the electric potential $\boldsymbol{\phi}_I=\begin{bmatrix} 1 & 0 &0 \end{bmatrix}$ for nodes $I=1,3,5,...$ and $ \boldsymbol{\phi}_I=\begin{bmatrix} 1/4 & 0 &0 \end{bmatrix}$ for nodes $I=2,4,6,...$ are applied in the initialization, the optimal control solution in Fig.~\ref{w-bc}(a) is obtained. When a mechanical configuration of bending and the electric potential $\boldsymbol{\phi}_I=\begin{bmatrix} 1 & 0 &0 \end{bmatrix}$ for nodes $I=1,3,5,...$ and $ \boldsymbol{\phi}_I=\begin{bmatrix} 1 & 0 &-1 \end{bmatrix}$ for nodes $I=2,4,6,...$ are applied in the initialization, the optimal control solution in Fig.~\ref{w-bc}(b) is obtained. The Lagrange multipliers and the electric charges are always initialized with zeros. The final position of the left cube is set as $x^{rb1}_N=2mm$ in the constraints for both cases. The motion of the left rigid cube is compared in Fig.~\ref{w-bc}(c). It can be observed that the final velocity in bending mode is higher than that in contraction. Larger differences of the motion can be expected when there is large difference of stiffness between bending and contraction.

\begin{figure}[htb!]
	\centering
	\subfigure[$t=0ms$]{\label{.}\includegraphics[width=.24\textwidth]{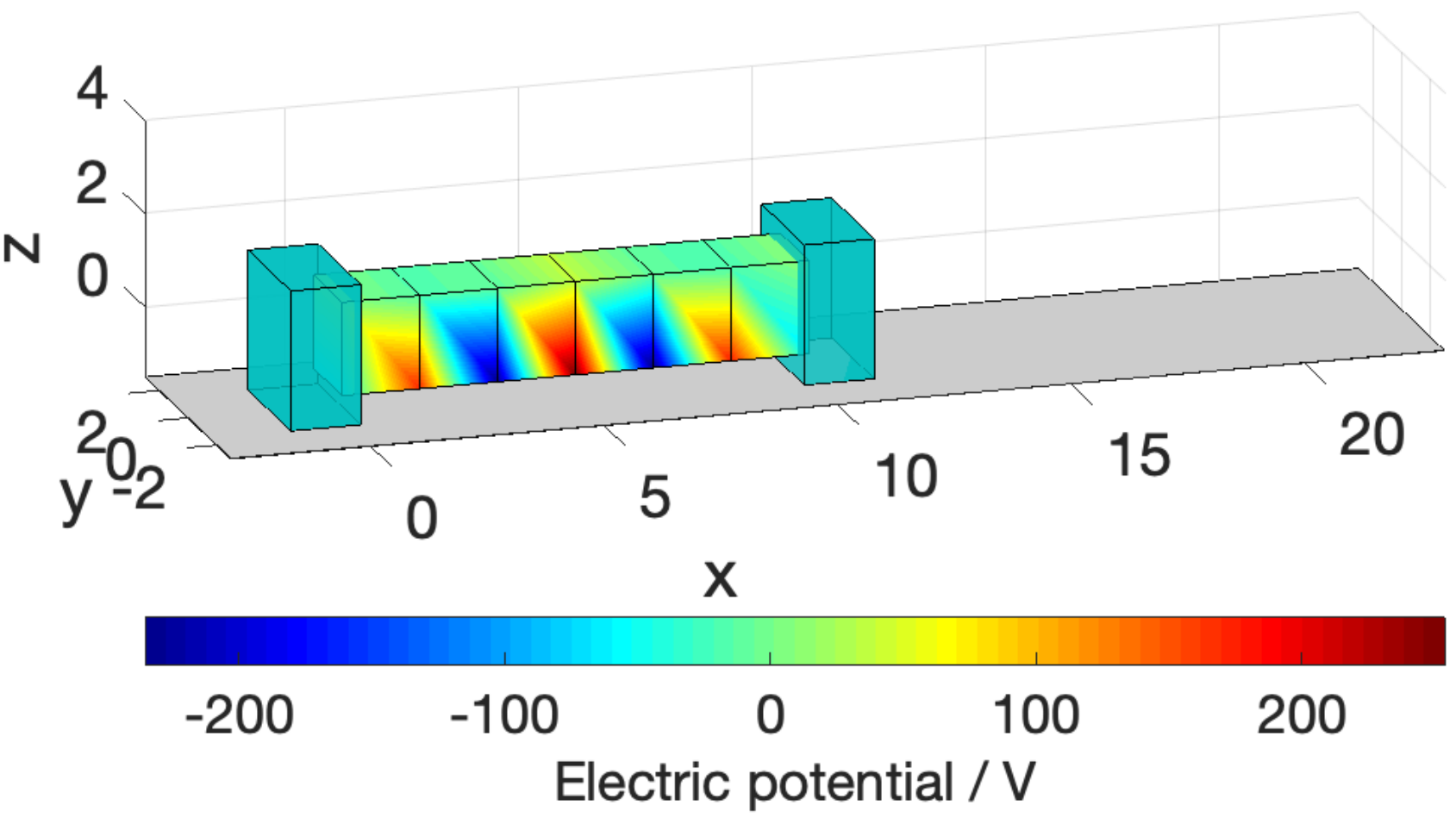}}
	\subfigure[$t=2ms$]{\label{.}\includegraphics[width=.24\textwidth]{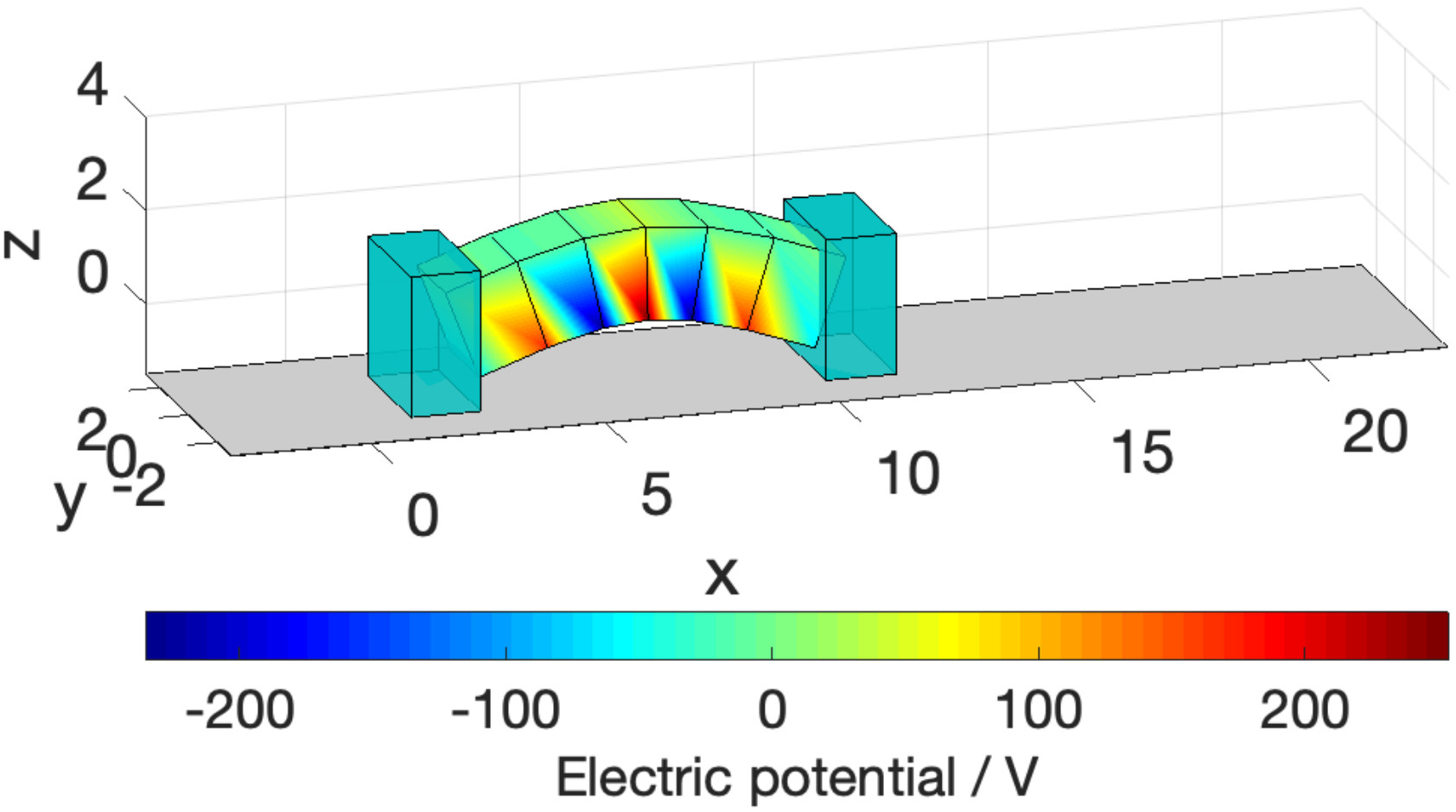}}
	\subfigure[$t=2.04ms$]{\label{.}\includegraphics[width=.24\textwidth]{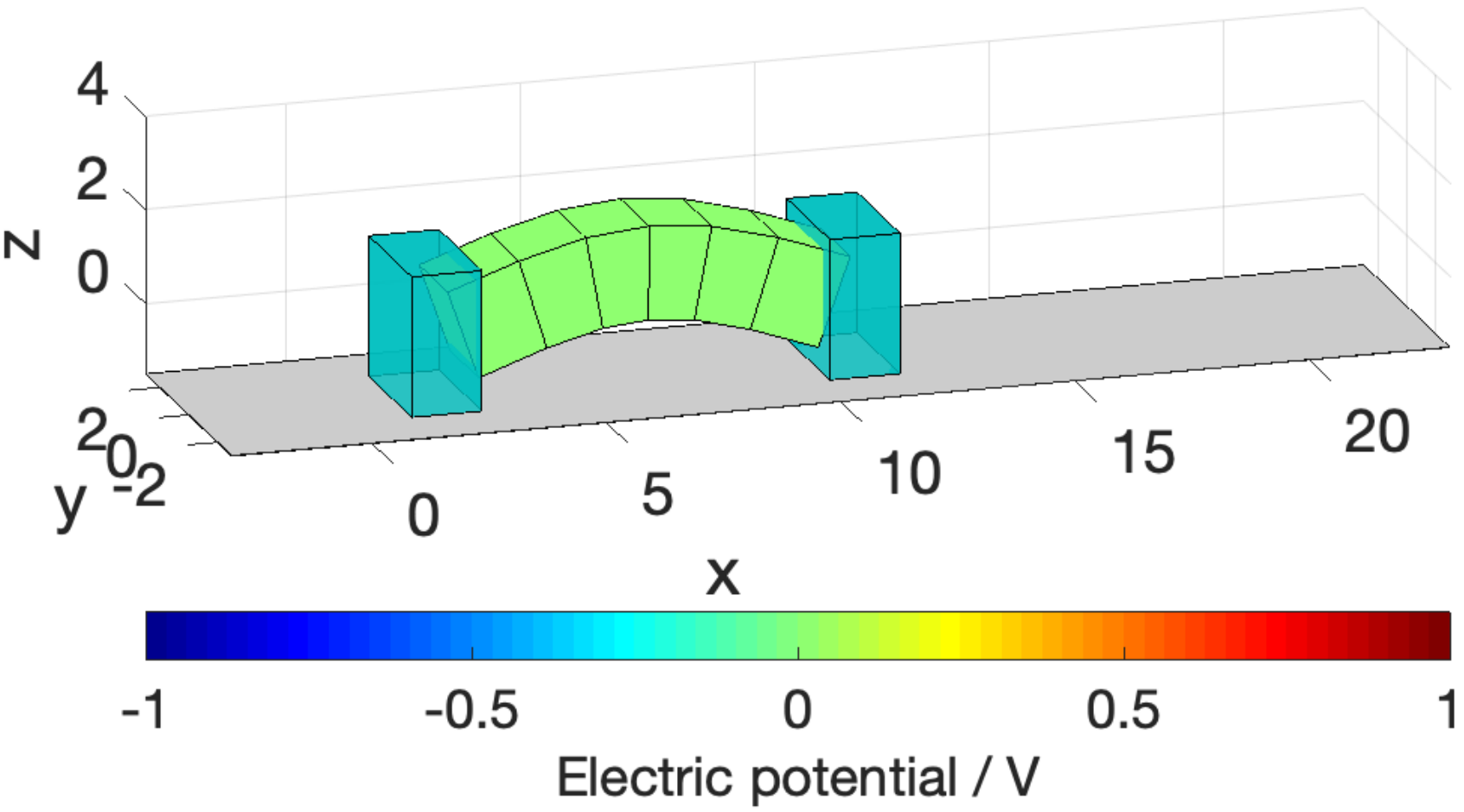}}
	\subfigure[$t=6ms$]{\label{.}\includegraphics[width=.24\textwidth]{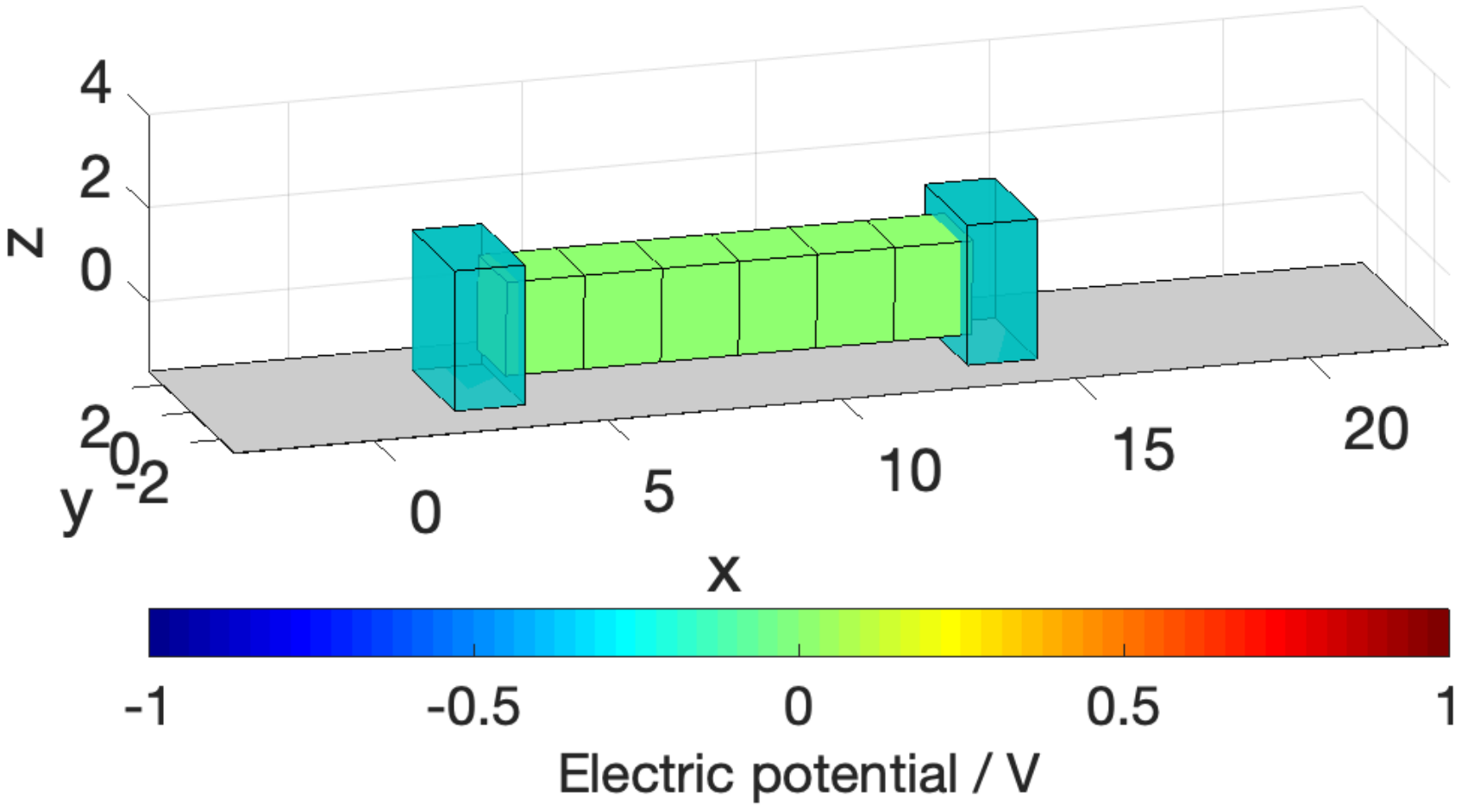}}\\
	\subfigure[$t=6.04ms$]{\label{.}\includegraphics[width=.24\textwidth]{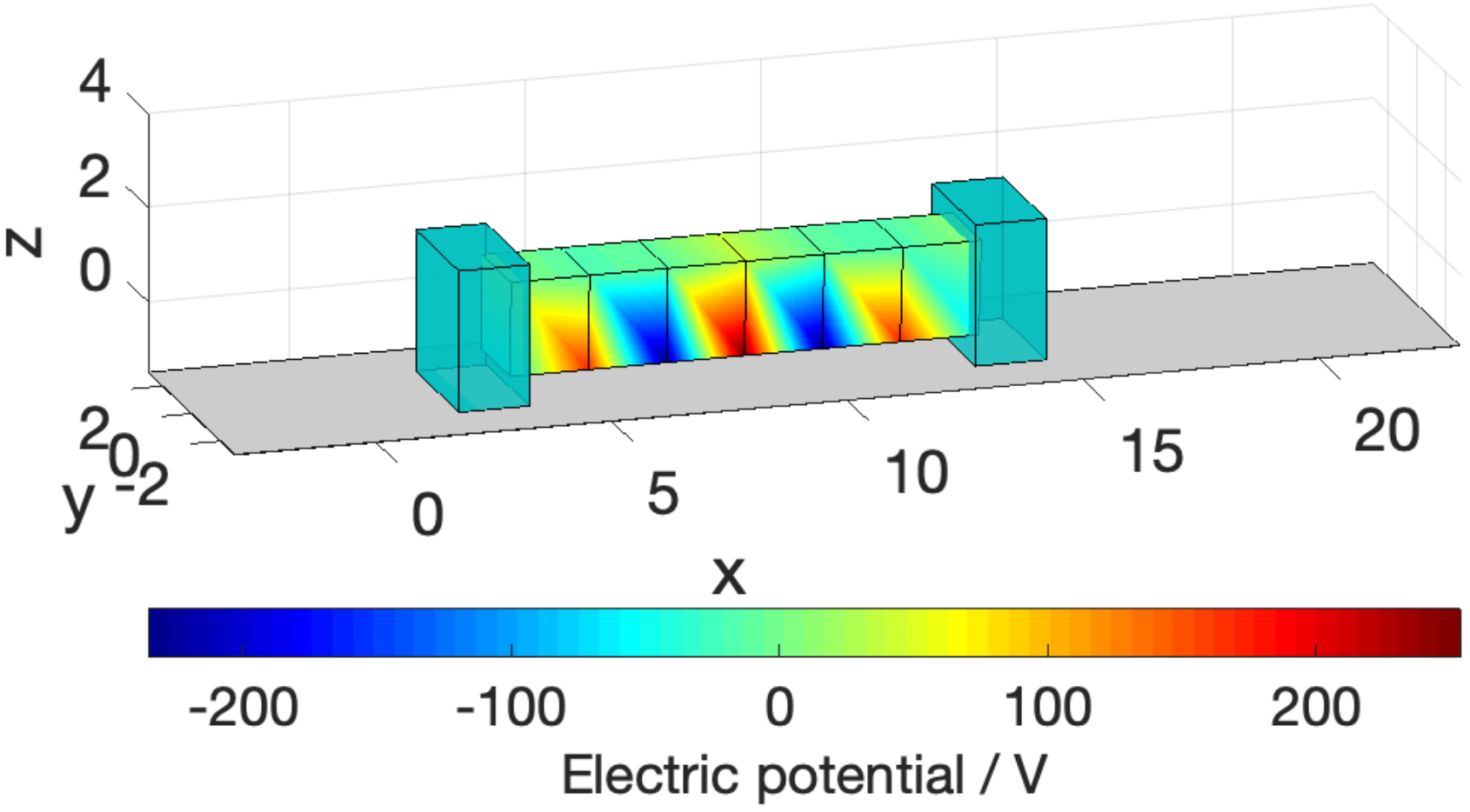}}
	\subfigure[$t=8ms$]{\label{.}\includegraphics[width=.24\textwidth]{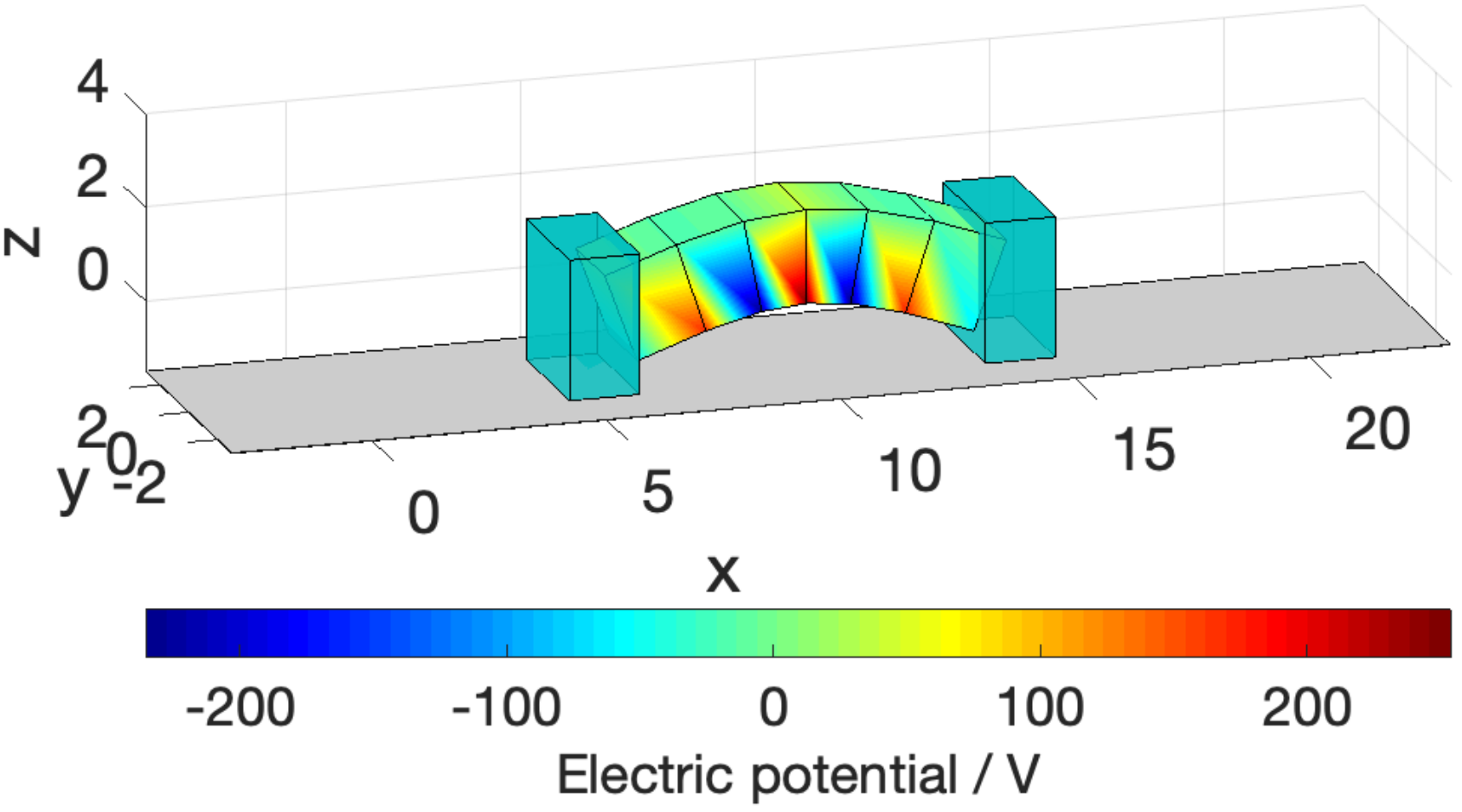}}
	\subfigure[$t=8.04ms$]{\label{.}\includegraphics[width=.24\textwidth]{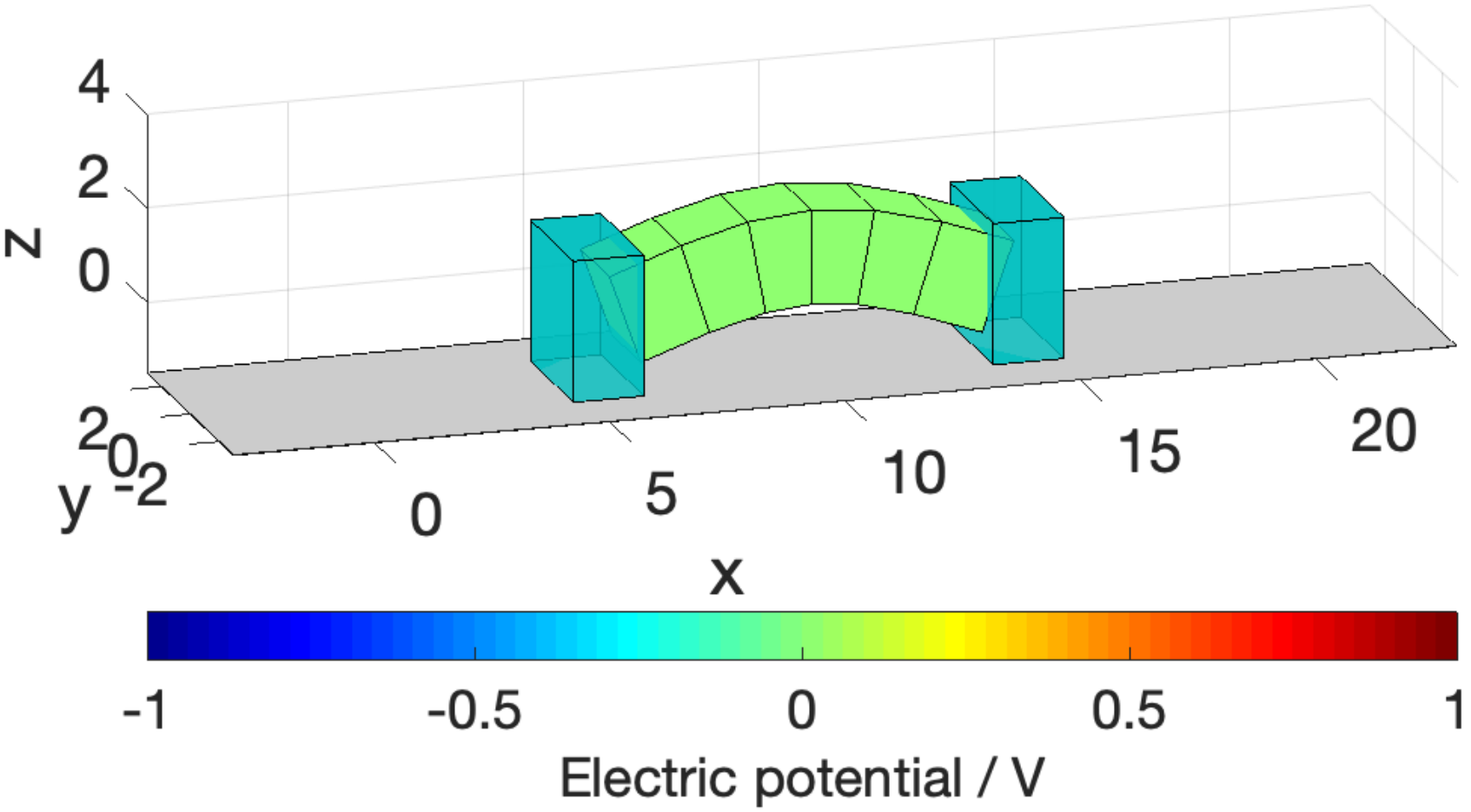}}
	\subfigure[$t=12ms$]{\label{.}\includegraphics[width=.24\textwidth]{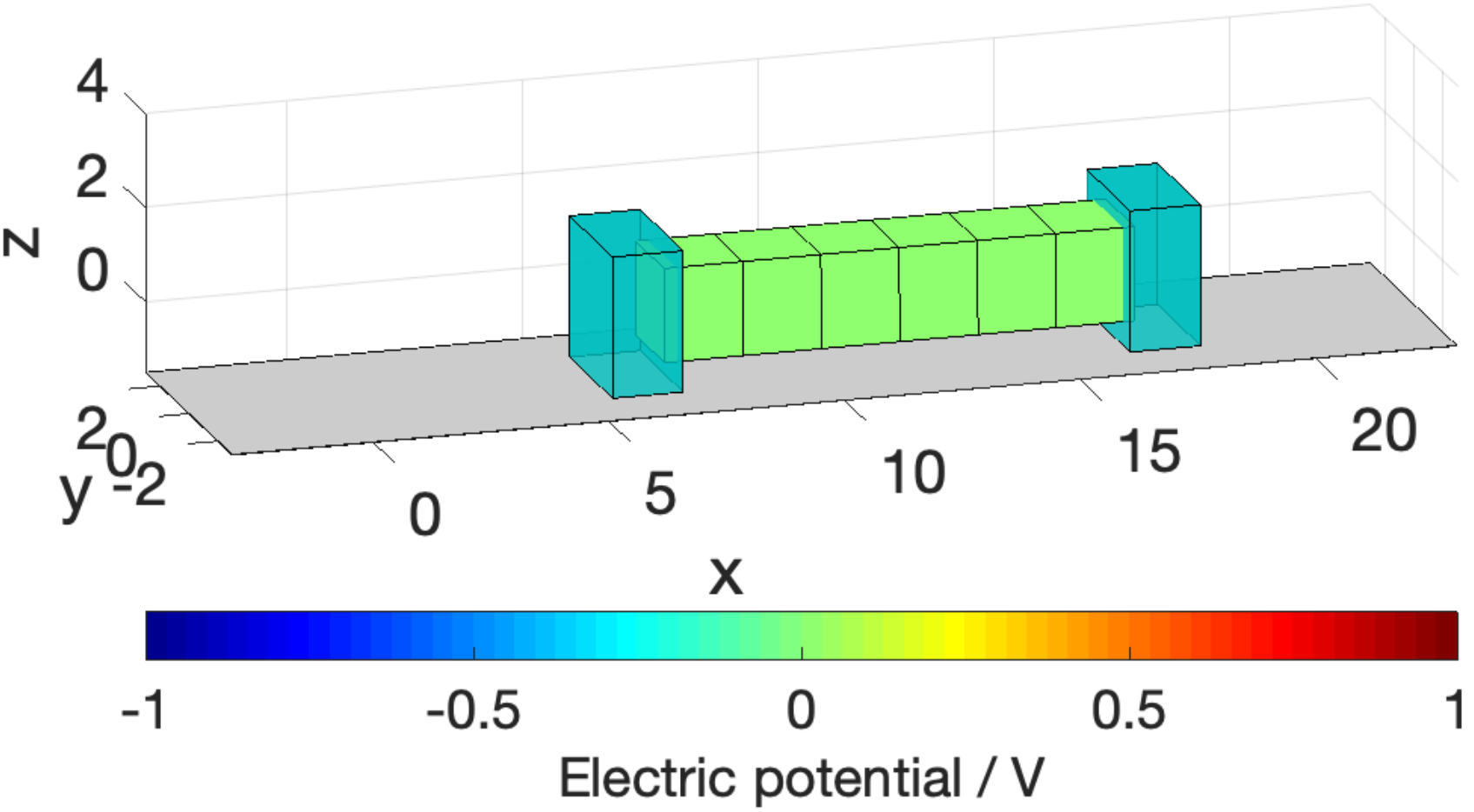}}
	
	\caption{Configurations of the  soft robotic worm.}
	\label{worm}
\end{figure}

\begin{figure}[htb!]
	\centering
	\subfigure[]{\includegraphics[width=.4\textwidth]{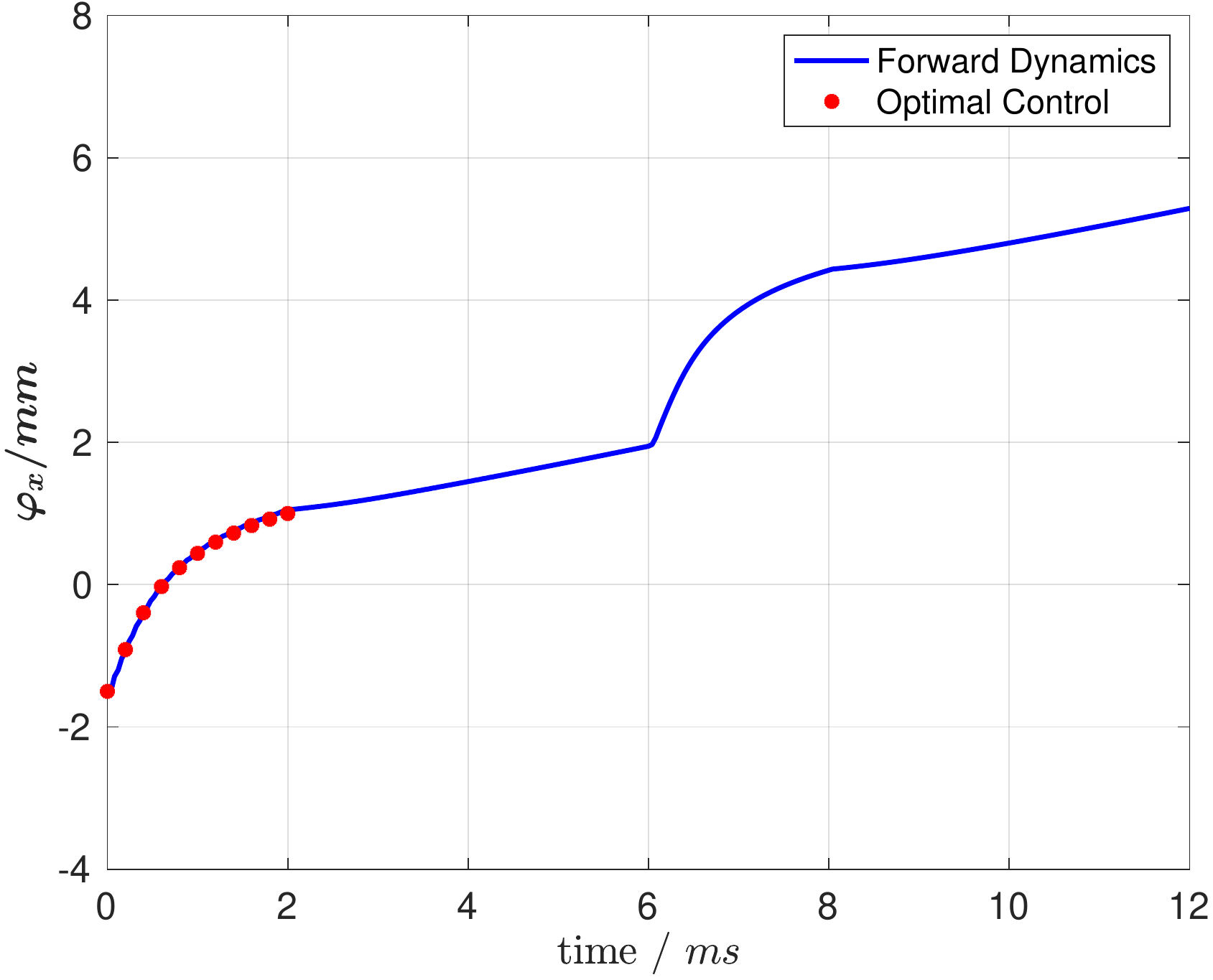}}
	\qquad
	\subfigure[]{\includegraphics[width=.4\textwidth]{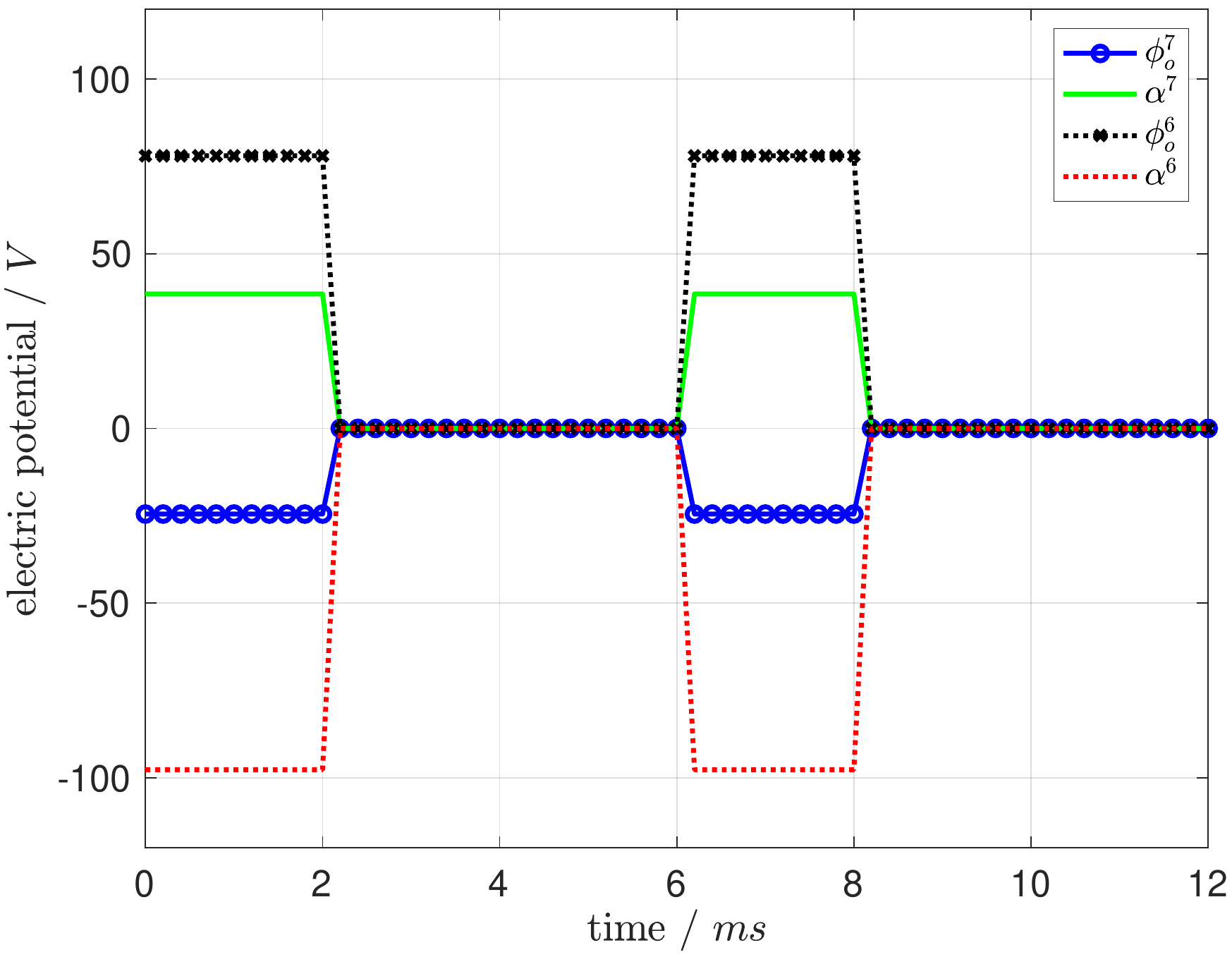}}
	\caption{(a) motion of the left rigid body and (b) electric potential of the last two beam nodes.}
	\label{worm-u}
\end{figure}

Next, the optimal trajectories of the electric potential and the electric charge in the bending mode are applied as electrical inputs for the forward dynamics simulation of a multiphase motion. In the stretching phase, the electric potential is set to zero. Fig.~\ref{worm} shows two cycles of actuated and stretching phases of the forward dynamics simulation. In the actuated phase, see Fig.~\ref{worm}(a) and (b), the bending is actuated by the electrical inputs. In the stretching phase, see Fig.~\ref{worm}(c) and (d), the free stretching of the beam is induced by the stored elastic energy from the previous phase. Fig.~\ref{worm}(e) to (h) show the second cycle of the two-phase motion with the same boundary conditions as the first cycle such that the forward motion can be observed. The motion of the left rigid cube is shown in Fig.~\ref{worm-u}(a), where the agreement of the optimal control trajectory in the actuated phase with the forward dynamics simulation can be observed. The left rigid cube moves forward during the stretching phase as well since its velocity is not zero at the end of the previous phase. Fig.~\ref{worm-u}(b) shows that the electric potentials are ensured to be constant in the actuated phases and go to zero in the stretching phases.

\subsection{Soft robotic grasper}
In the last example, the optimal control of a soft robotic two-finger grasping problem is investigated, where two electromechanically coupled beams represent soft robotic fingers that grasp a rigid cylinder. Here, the contact between the beam actuators and the rigid body in a flexible multibody system has to be considered. When bending of the beams is actuated, the rigid cylinder can be held via the unilateral contact constraints. The geometry of the longer beam in reference configuration is the same as in \mbox{Section 5.1}. The longer beam is discretized with 10 DEA cells with one element in each. The geometry of the shorter beam in reference configuration is given as: length $l=6mm$ and width $b=2mm$. The shorter beam is discretized with 6 DEA cells with one element in each. The rigid cylinder with radius $R_{\rm rb}=2mm$ and mass $m_{\rm rb}=0.5$ is placed at $(5,3.1,0)$ at $t=0ms$. The placement of the cylinder in $z$-direction is fixed. The frictionless contact between the beams and the cylinder is described with the following normal gap function
\begin{align}
	g^{c,I}=\left\| \left( \boldsymbol{\varphi}_x^I, \boldsymbol{\varphi}_y^I\right) -\left( \boldsymbol{\varphi}_x^{\rm rb} , \boldsymbol{\varphi}_y^{\rm rb}\right) \right\| - R_{rb} \geqslant 0,
\end{align}
where $(\boldsymbol{\varphi}^I_x, \boldsymbol{\varphi}^I_y)$ and $(\boldsymbol{\varphi}^{\rm rb}_x, \boldsymbol{\varphi}^{\rm rb}_y)$ are the transnational configurations of the beam node $I$ and the cylinder in the $(x,y)$-plane, respectively.

In the initial state, both of the beams are straight and the cylinder is placed between them without any contact. This mechanical configuration of the multibody system is applied as the initial condition in the equality constraints in Eq.~(\ref{ini}). In the final condition in Eq.~(\ref{final}), only the final position of the cylinder is constrained whereas the final condition for the beams are released, which leads to a grasp in which the optimization determines where the contact shall be closed. The time step is set as $\Delta t=0.04ms$ and the total time is set as $t=0.8ms$. The damping parameter is set as $\eta=15$. 

Before solving the optimization problem, the optimization variables are initialized as following. The mechanical trajectory is initialized with a linear distribution between the initial mechanical configuration $\mathbf{q}_0^{\rm mech}$ and a bending state $\mathbf{q}_N^{\rm mech}$. The electrical potentials are initialized with constant values $\boldsymbol{\phi}_I=\begin{bmatrix} 1 & 0 &0 \end{bmatrix}$ for nodes $I=1,3,5,...$ and $ \boldsymbol{\phi}_I=\begin{bmatrix} 1 & 1 &0 \end{bmatrix}$ for nodes $I=2,4,6,...$. The Lagrange multipliers and the electric charges are initialized with zeros.

\begin{figure}[htb!]
	\centering
	\subfigure[$t=0ms$]{\label{.}\includegraphics[width=.32\textwidth]{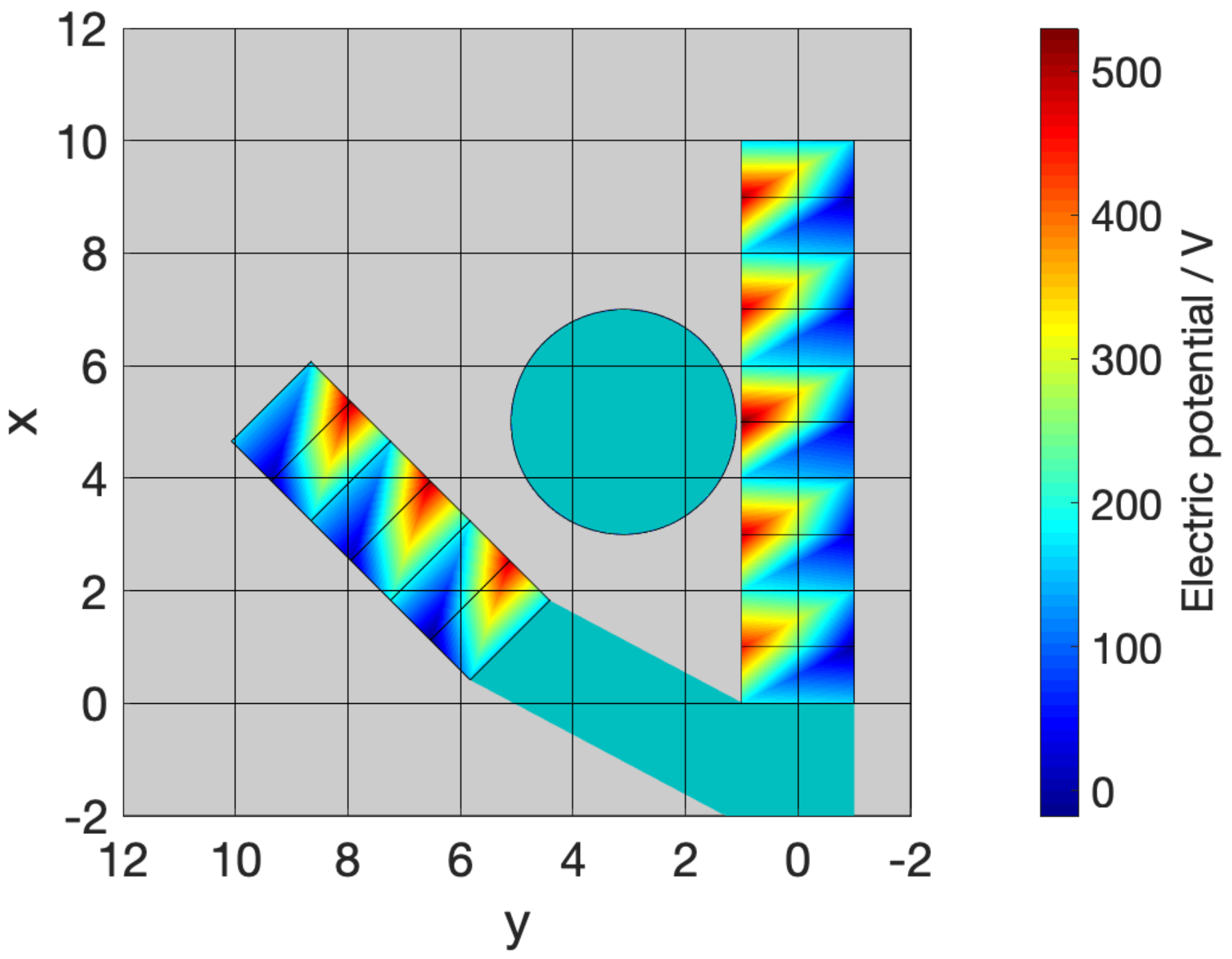}}
	\subfigure[$t=0.8ms$]{\label{.}\includegraphics[width=.32\textwidth]{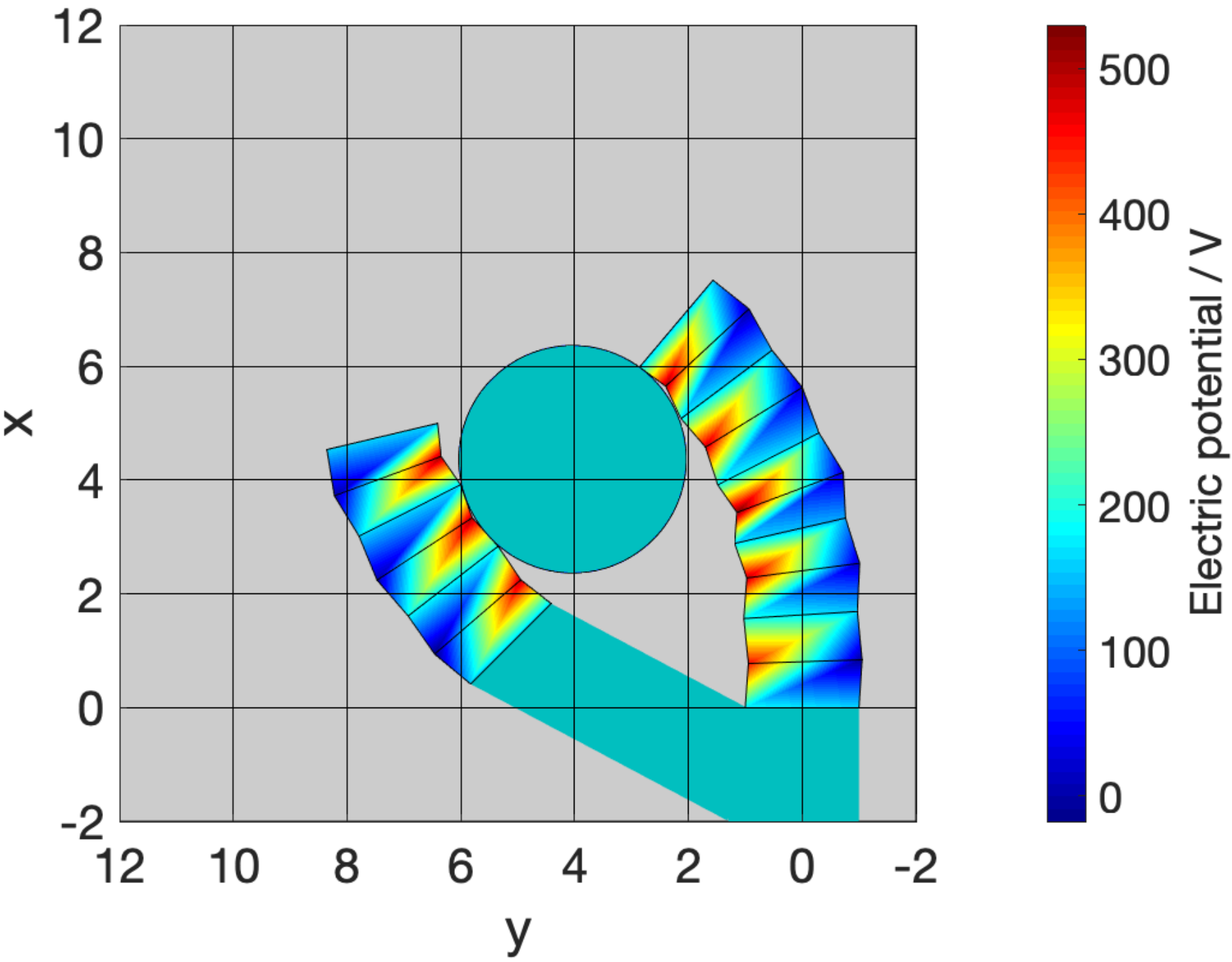}}
	\subfigure[$t=0.8ms, 3D \text{ view}$]{\label{.}\includegraphics[width=.32\textwidth]{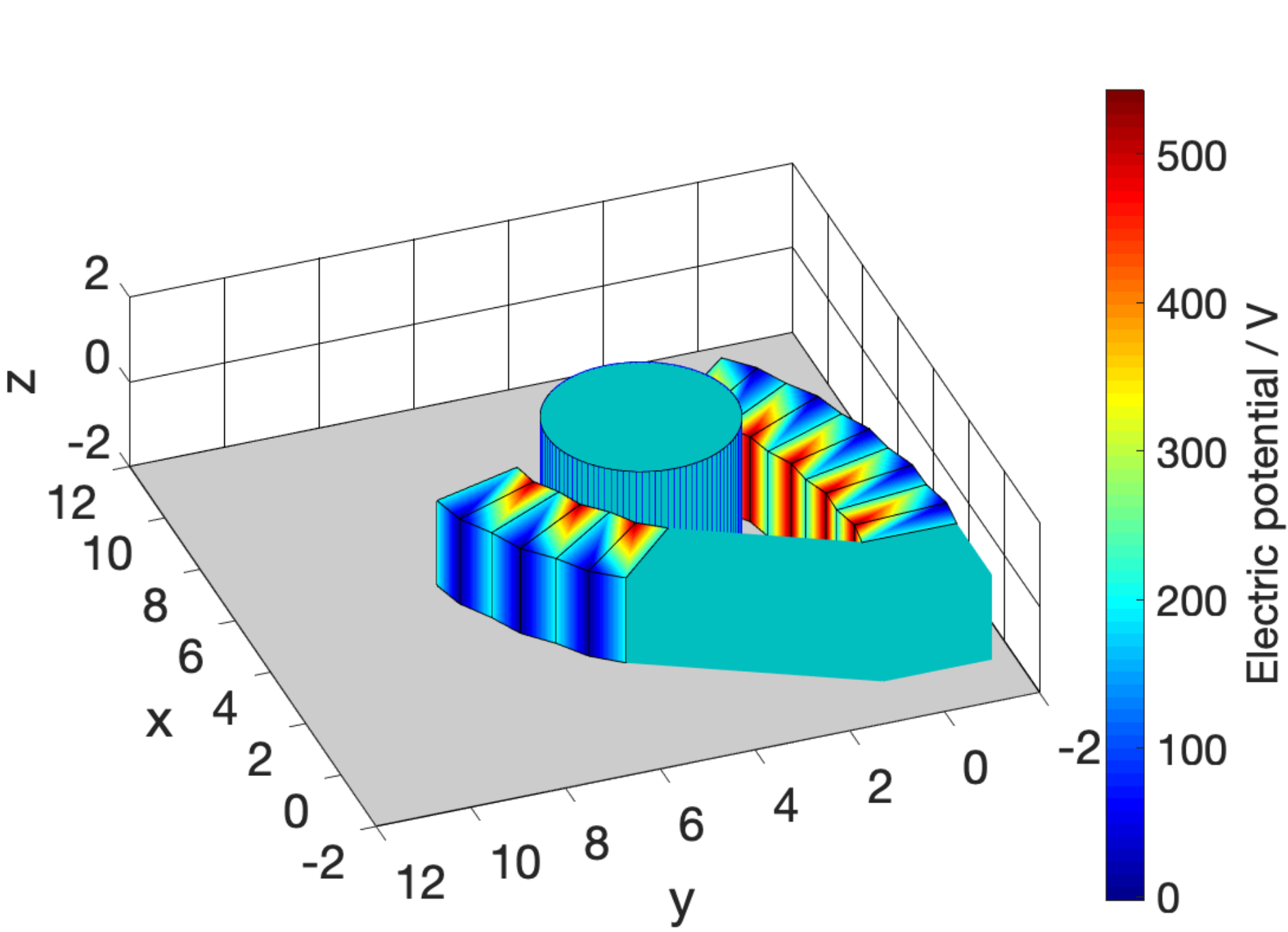}}
	\caption{Configurations of the soft robotic grasper.}
	\label{arm}
\end{figure}

\begin{figure}[htb!]
	\centering
	\subfigure[longer beam]{\label{.}\includegraphics[width=.32\textwidth]{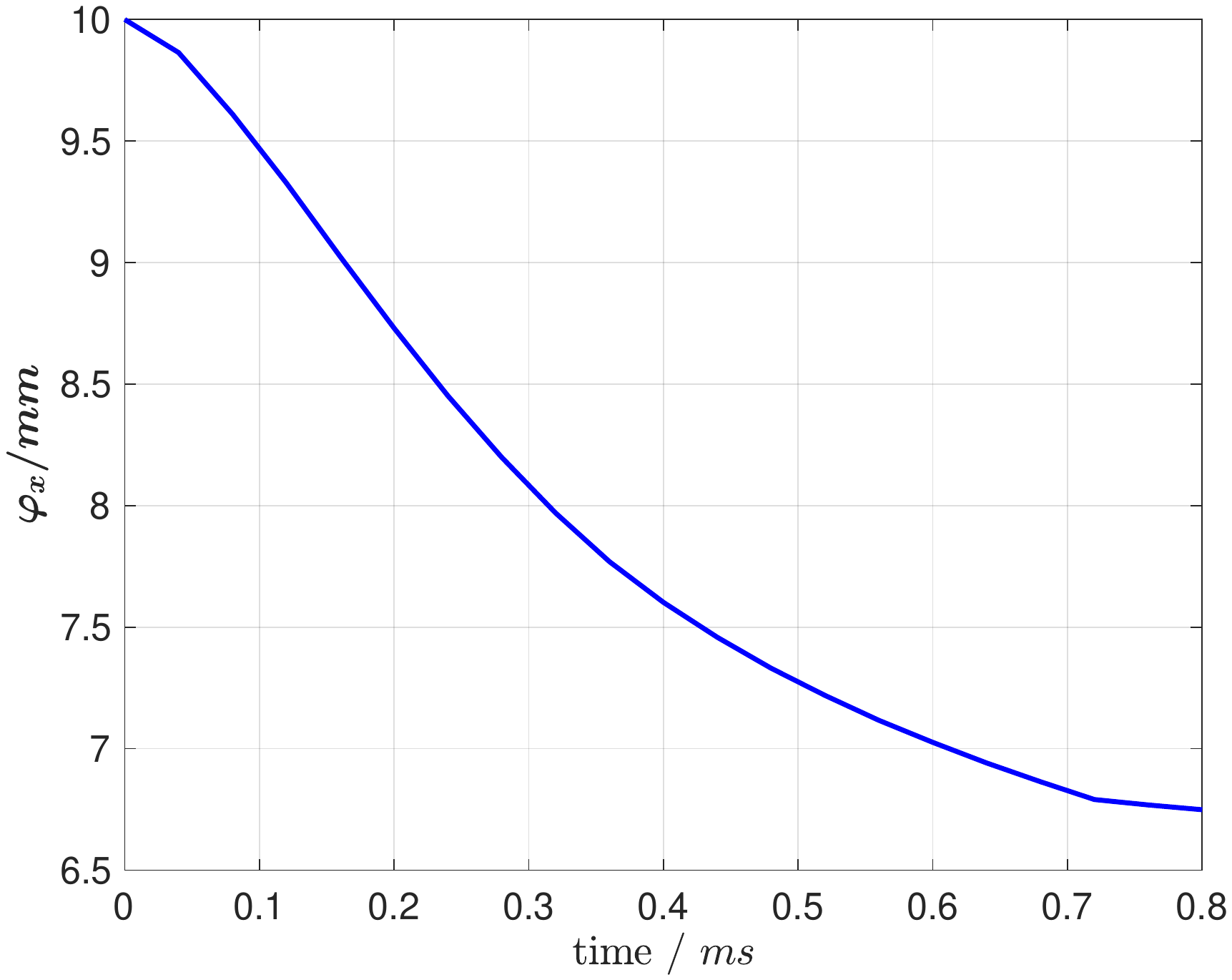}}
	\subfigure[shorter beam]{\label{.}\includegraphics[width=.32\textwidth]{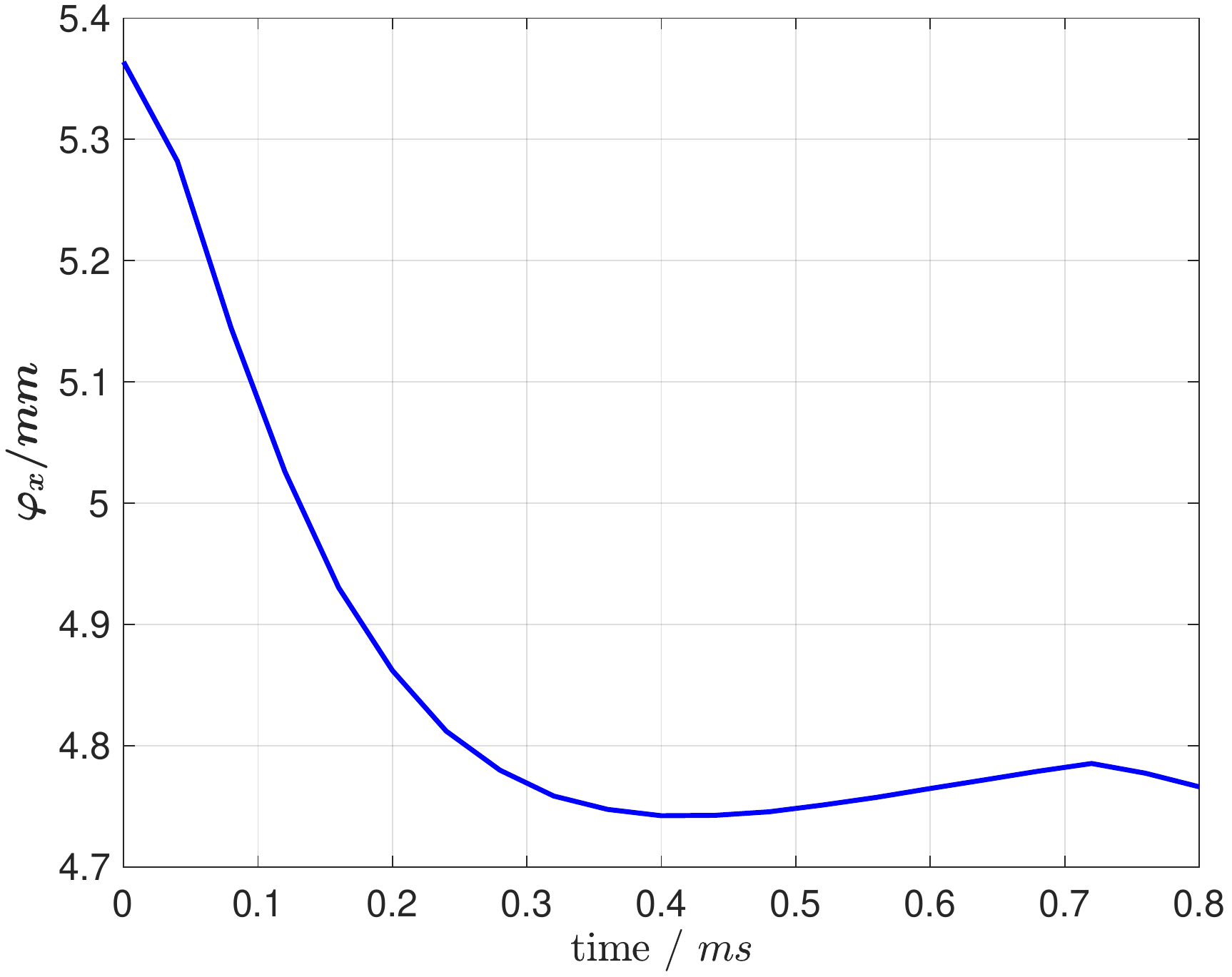}}
	\subfigure[rigid cylinder]{\label{.}\includegraphics[width=.32\textwidth]{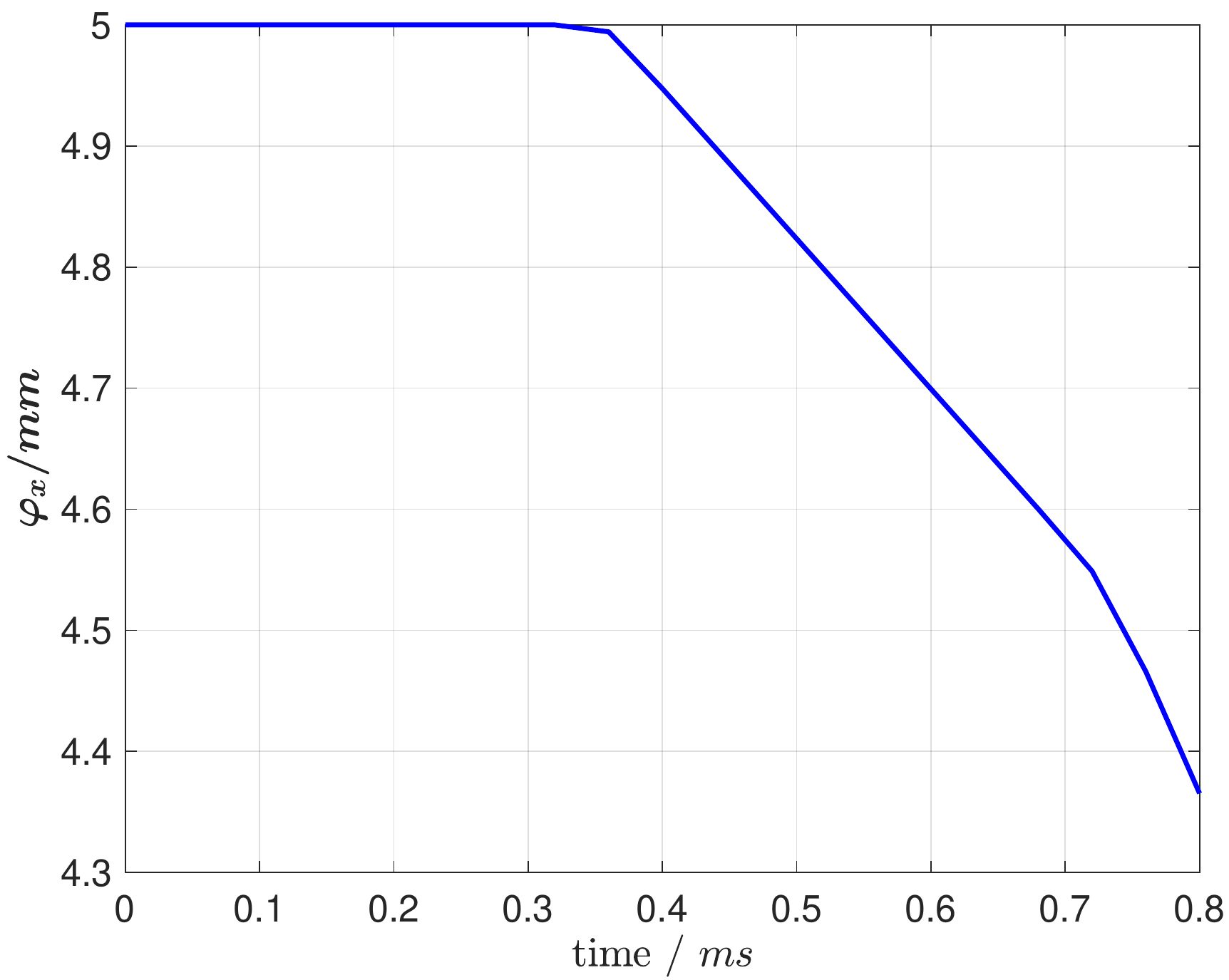}}\\
	\subfigure[longer beam]{\label{.}\includegraphics[width=.32\textwidth]{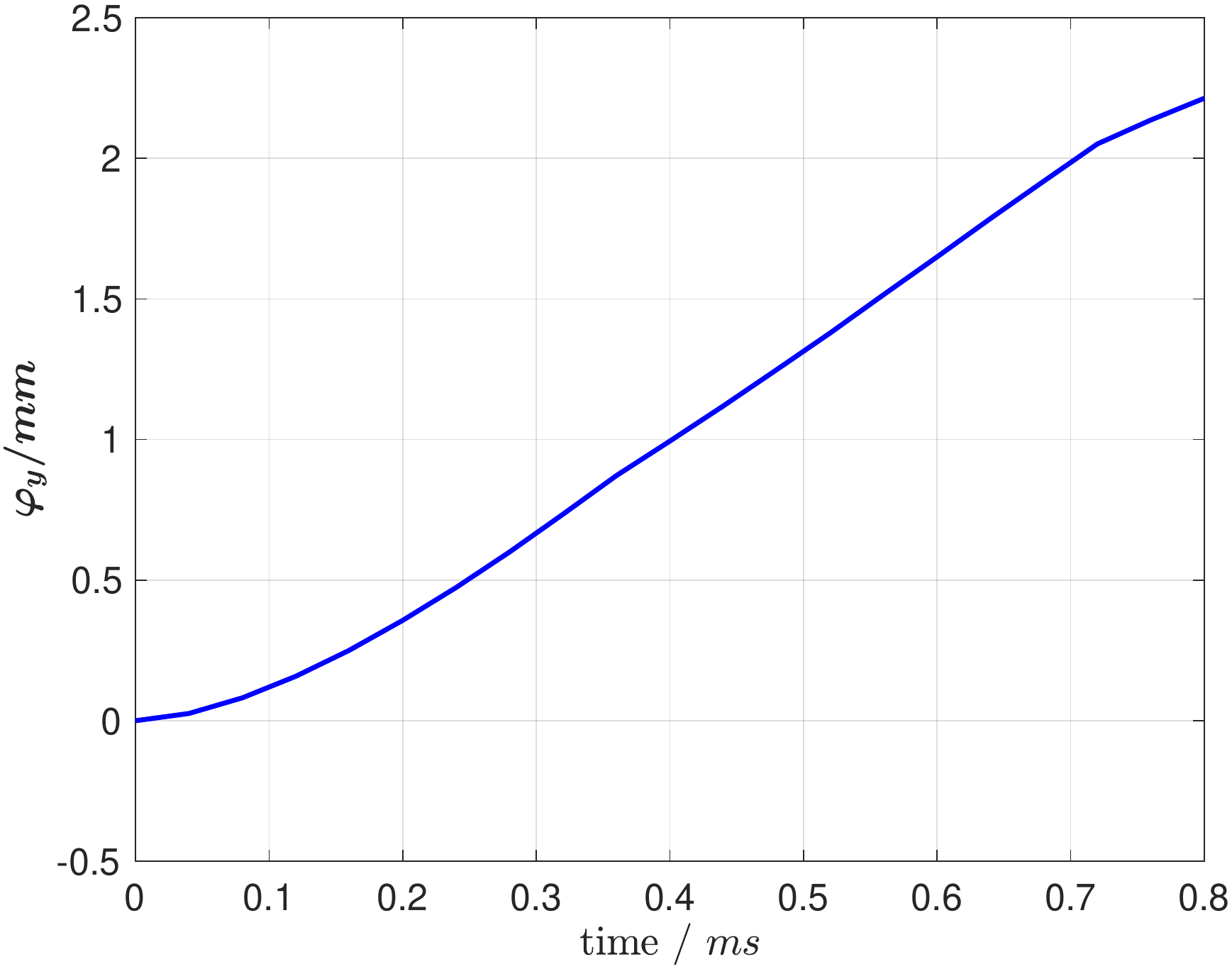}}
	\subfigure[shorter beam]{\label{.}\includegraphics[width=.32\textwidth]{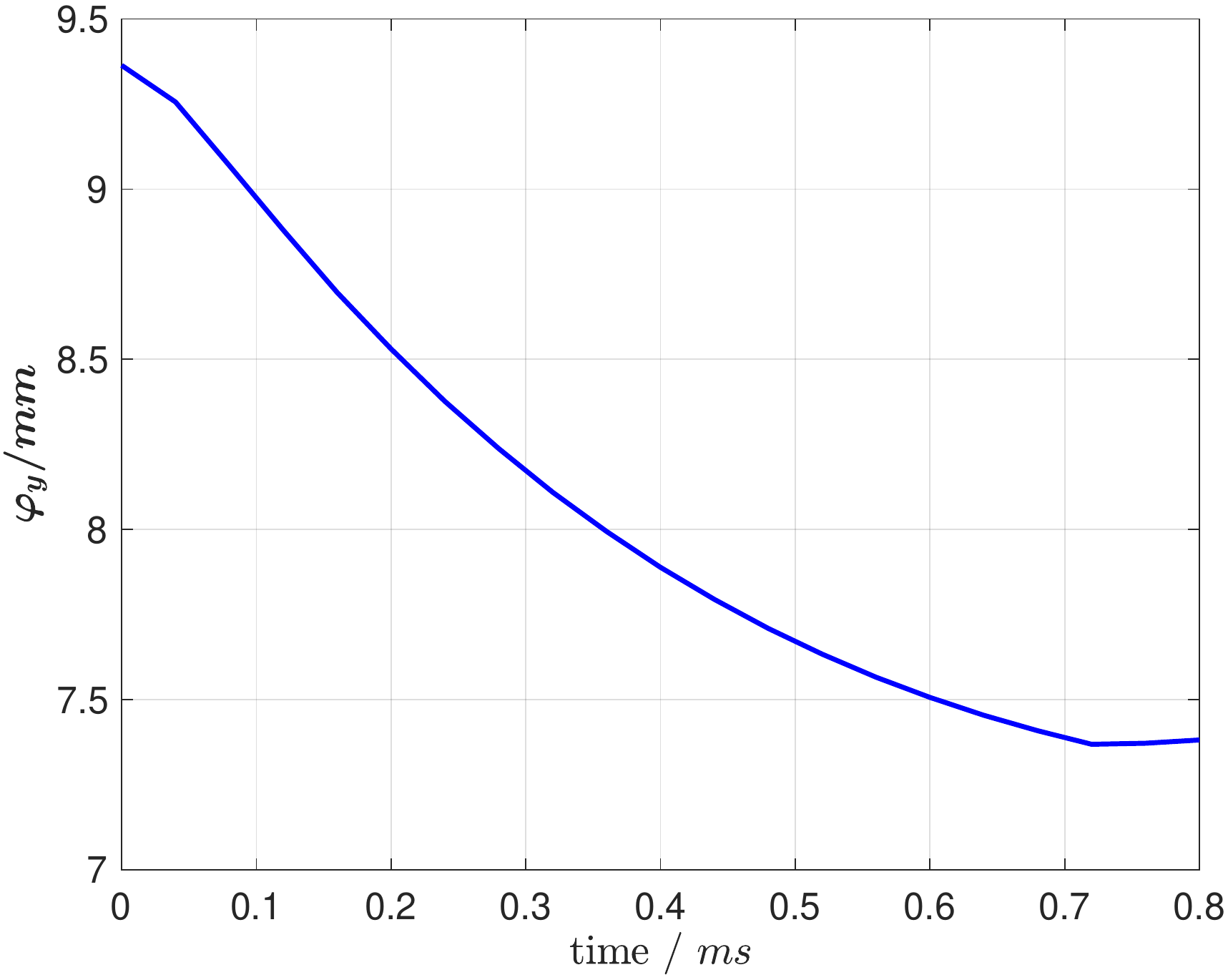}}
	\subfigure[rigid cylinder]{\label{.}\includegraphics[width=.32\textwidth]{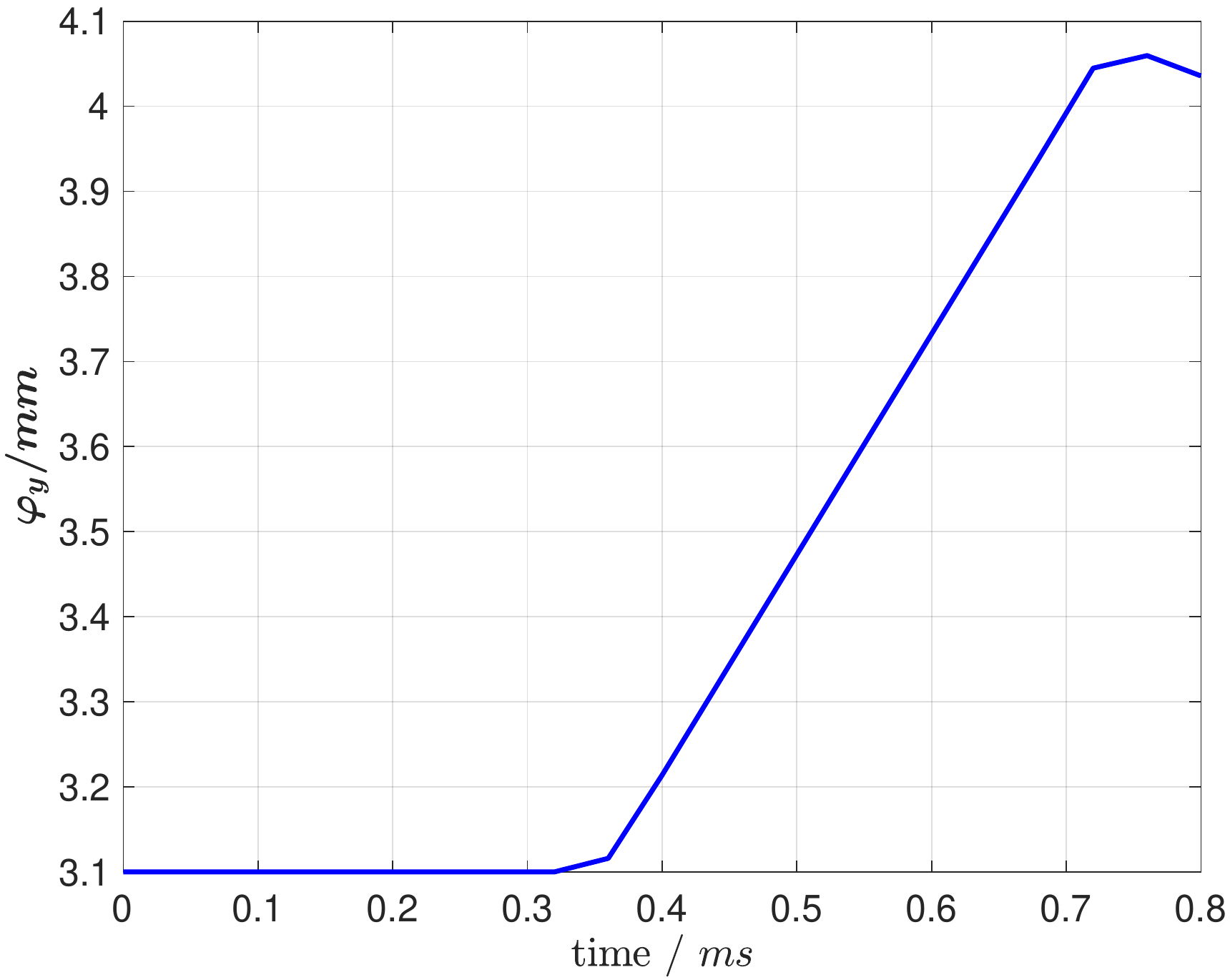}}
	\caption{Motion of the beam nodes at the free ends and the cylinder.}
	\label{arm-u}
\end{figure}

\begin{figure}[htb!]
	\centering
	\subfigure[longer beam]{\label{.}\includegraphics[width=.4\textwidth]{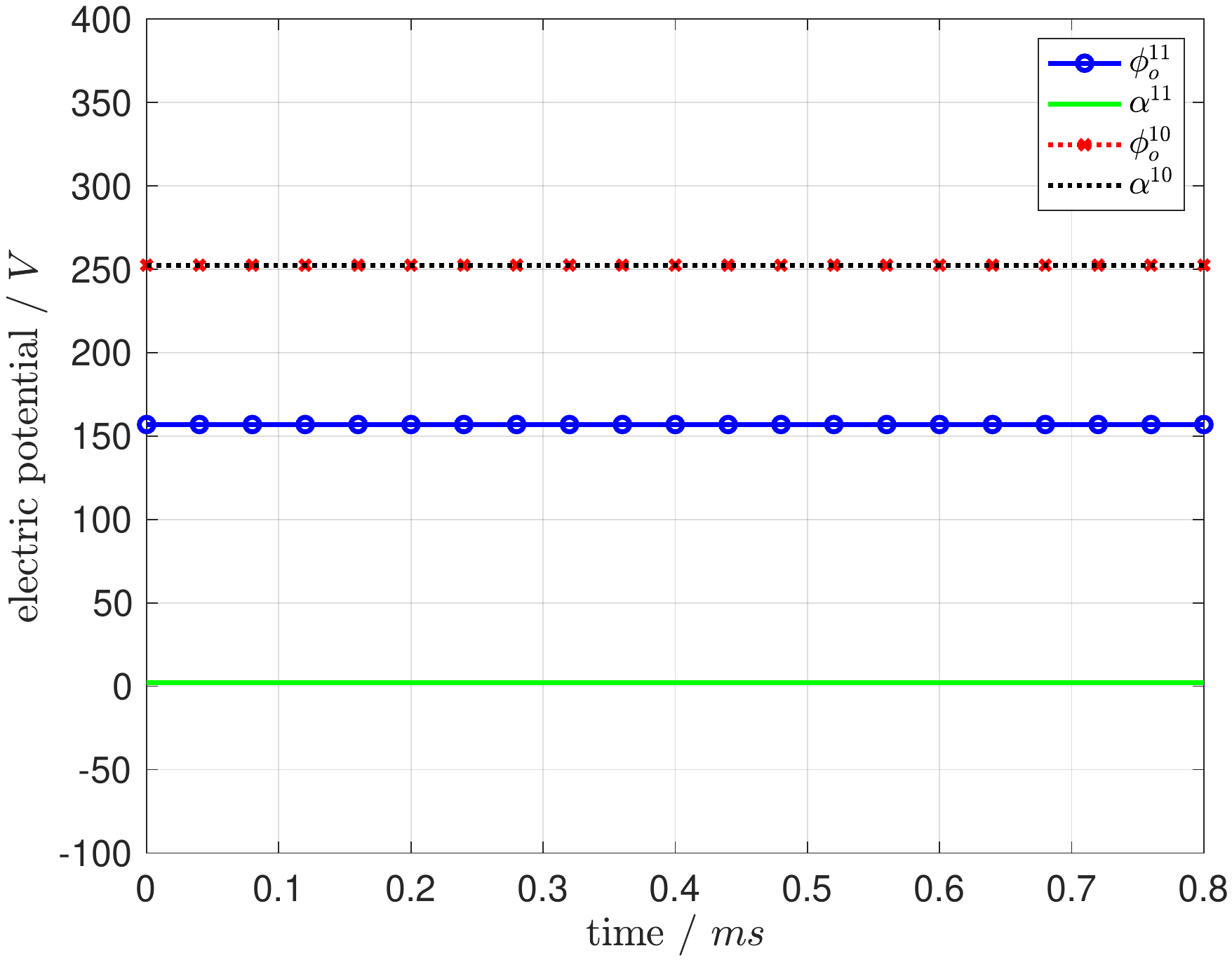}}
	\qquad
	\subfigure[shorter beam]{\label{.}\includegraphics[width=.4\textwidth]{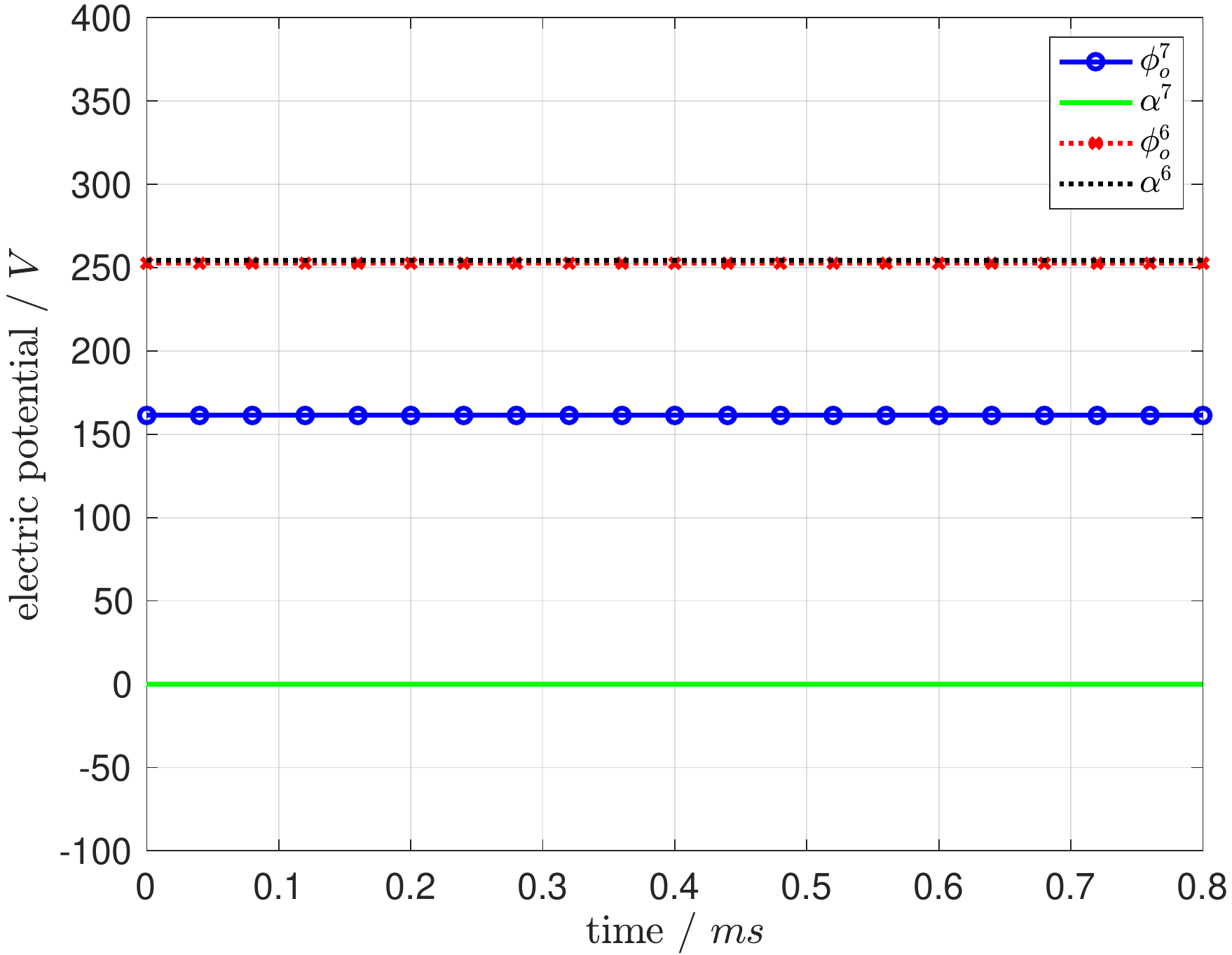}}
	\caption{Electric potential of the beams.}
	\label{arm-v}
\end{figure}

\begin{figure}[htb!]
	\centering
	\includegraphics[width=.5\textwidth]{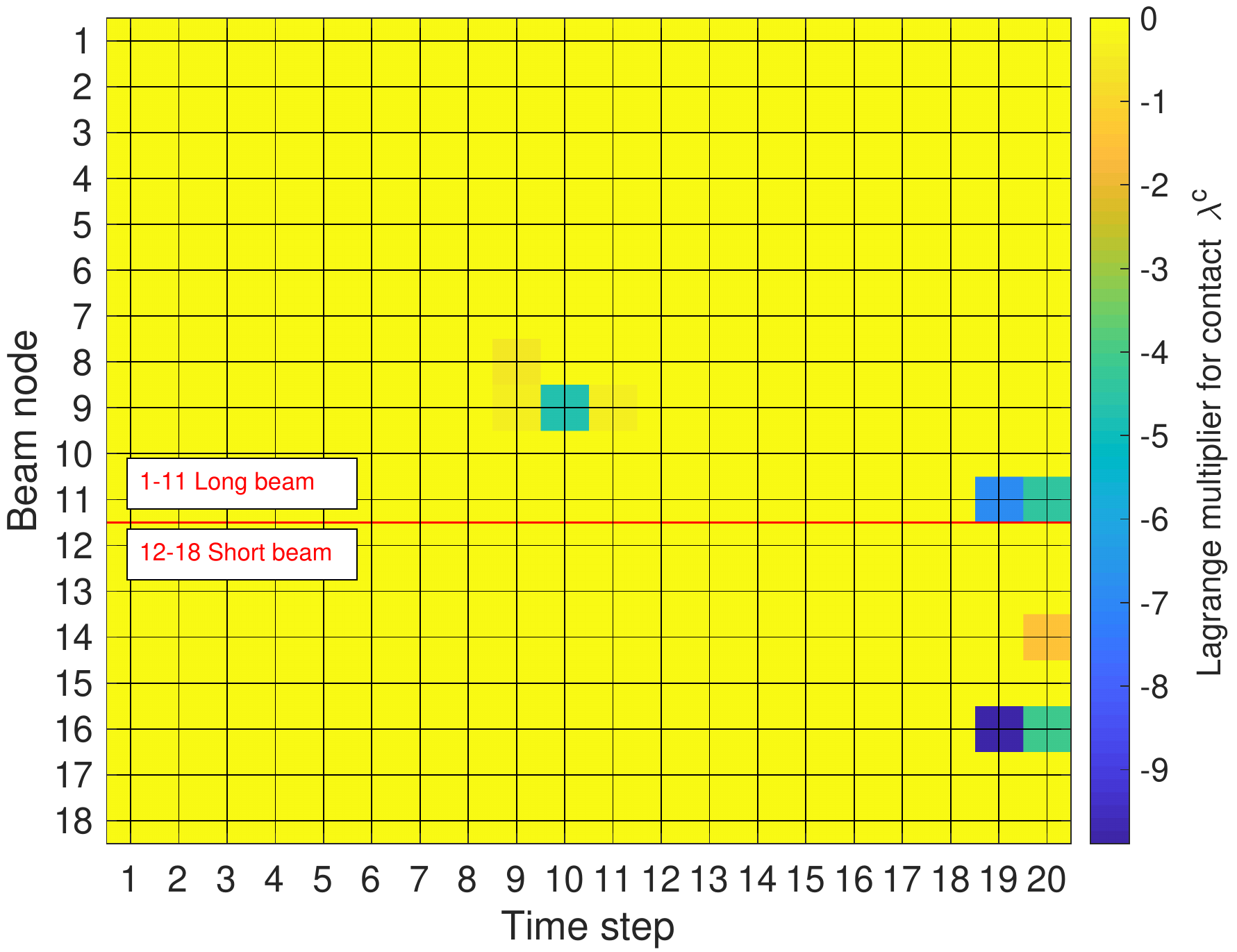}
	\caption{Contact forces.}
	\label{a-ct}
\end{figure}

By solving the optimization problem, the optimal trajectories are obtained as shown in Fig.~\ref{arm}, where the initial and final configurations are plotted. By means of the bending and contraction actuated in the beams, the rigid cylinder is firstly moved and finally held at the desired position via its contacts with the beams. The motion of the rigid cylinder and the beam nodes at the free ends is shown in Fig.~\ref{arm-u}. The electrical trajectories of the last two beam nodes are shown in Fig.~\ref{arm-v}, where the constant electric potential over time can be observed. Fig.~\ref{a-ct} shows the trajectories of the Lagrange multiplier $\lambda^c$ for contact, where the contact constraints are closed when the value of $\lambda^c$ is nonzero. From Fig.~\ref{a-ct}, it can be seen that the rigid cylinder contacts with the nodes 8 and 9 of the long beam firstly. Then the contact is open for some time steps, which can be understood as the approaching phase. From the time step 19, the cylinder is held by the two beams with 2 to 3 contact pairs, which can be understood as the holding phase of the robotic grasping problem.

\section{Conclusion}
In this paper, the optimal control problem of the electromechanically coupled flexible multibody dynamics system is formulated and solved. The mathematically concise and physically representative free energy function together with the work conjugated beam variables works well in the flexible mutibody dynamics systems in this work. The constrained discrete Euler-Lagrange equations of the coupled multibody dynamics system complemented with the Kuhn-Tucker conditions for contacts are applied as constraints in the optimal control formulation. By treating the electric potential on the beam nodes as control objective, the optimal trajectories of the configurations, the Lagrange multipliers and the electric charges are found after solving the optimization problem. In the numerical examples, the minimum change of the electric potential on the beam nodes are ensured, which brings conveniences for the power supply. The constraints of the optimization problem are fulfilled exactly in the numerical results, such as the dielectric elastomer actuated bending of a beam and the dielectric elastomer actuated locomotion of a soft robotic worm. In particular, the optimal control formulation works for a two-finger soft robotic grasper, where the soft robotic grasping based on contact is accomplished by imposing inequality constraints in the optimization.

\section*{Acknowledgements}
The authors acknowledge the support of Deutsche Forschungsgemeinschaft (DFG) with the project: LE 1841/7-1.\\

\vspace*{20\baselineskip}
\section*{Appendix}
The derivation of the reduced strain energy in Eq.~(\ref{redu}) for the beam is conducted as following:
\begin{align}
	\Omega^r_e(\boldsymbol{\Gamma}, \mathbf{K}, \boldsymbol{\Xi}, \boldsymbol{\Theta})&=c_1 \mathbf{E}^e \cdot \mathbf{E}^e + c_2 \mathbf{C}^{\rm linear}:(\mathbf{E}^e \otimes \mathbf{E}^e)\\
	&\nonumber\\
	&=(c_1 +c_2) \left[  \Xi_1^2 +  \Xi_2^2 + ( \Xi_3+ \Theta_1X^1 + \Theta_2X^2)^2  \right] \nonumber\\
	&\;\;\;\;\; + 2c_2 (\Xi_3 + \boldsymbol\Theta\cdot\mathbf{X}) \left[  \Xi_1(\mathbf{d}_1^0 \cdot  \mathbf{a}^r  ) +\Xi_2(\mathbf{d}_2^0 \cdot \mathbf{a}^r ) +(\Xi_3 + \boldsymbol\Theta\cdot\mathbf{X})(\mathbf{d}_3^0 \cdot \mathbf{a}^r)  \right]\\
	&\nonumber\\
	& = (c_1 +c_2) \left[  \Xi_1^2 +  \Xi_2^2 + ( \Xi_3+ \Theta_1X^1 + \Theta_2X^2)^2  \right] \nonumber\\
	&\;\;\;\;\; + 2c_2  \Xi_3( \Xi_1 \mathbf{d}_1^0 \cdot \mathbf{a}^r +\Xi_2 \mathbf{d}_2^0 \cdot \mathbf{a}^r +\Xi_3 \mathbf{d}_3^0 \cdot \mathbf{a}^r  ) +  \Xi_3 (\boldsymbol\Theta\cdot\mathbf{X})(\mathbf{d}_3^0 \cdot \mathbf{a}^r )  \nonumber \\
	& \;\;\;\;\; + 2c_2 \boldsymbol\Theta\cdot\mathbf{X} ( \Xi_1 \mathbf{d}_1^0 \cdot \mathbf{a}^r +\Xi_2 \mathbf{d}_2^0 \cdot \mathbf{a}^r +\Xi_3 \mathbf{d}_3^0 \cdot \mathbf{a}^r ) + (\boldsymbol\Theta\cdot\mathbf{X})^2(\mathbf{d}_3^0 \cdot \mathbf{a}^r )\\
	& \nonumber\\
	& = (c_1 +c_2) \left[  \Xi_1^2 +  \Xi_2^2 + ( \Xi_3+ \Theta_1X^1 + \Theta_2X^2)^2  \right] \nonumber\\
	&\;\;\;\;\; + 2c_2 \left[  \Xi_1 \Xi_3\Gamma_1 +\Xi_2 \Xi_3 \Gamma_2 +(\Xi_3^2 + \Theta_1^2 X_1^2 +\Theta_2^2 X_2^2) \Gamma_3 \right] \nonumber\\
	& \;\;\;\;\; + 2c_2 \boldsymbol\Theta\cdot\mathbf{X} \left[  \Xi_1 \mathbf{d}_1^0 \cdot (\mathbf{K} \times X_k  \mathbf{d}^0_k) +\Xi_2 \mathbf{d}_2^0 \cdot (\mathbf{K} \times X_k  \mathbf{d}^0_k)+2\Xi_3 \mathbf{d}_3^0 \cdot (\mathbf{K} \times X_k  \mathbf{d}^0_k)\right] \\
	& \nonumber\\
	& = (c_1 +c_2) \left[  \Xi_1^2 +  \Xi_2^2 + ( \Xi_3+ \Theta_1X^1 + \Theta_2X^2)^2  \right] \nonumber\\
	&\;\;\;\;\; + 2c_2 \left[  \Xi_1 \Xi_3\Gamma_1 +\Xi_2 \Xi_3 \Gamma_2 +(\Xi_3^2 + \Theta_1^2 X_1^2 +\Theta_2^2 X_2^2) \Gamma_3 \right] \nonumber\\
	& \;\;\;\;\; + 2c_2 (-\Xi_1\Theta_2K_3 X_2^2  + \Xi_2\Theta_1K_3X_1^2 + 2\Xi_3 \Theta_2K_1X_2^2 -2\Xi_3\Theta_1K_2X_1^2) \nonumber\\
	& \;\;\;\;\; + 2c_2 (-\Theta_1 \Xi_1 X_1 X_2 K_3 +2\Theta_1\Xi_3X_1X_2K_1+\Theta_2\Xi_2X_1X_2K_3-2\Theta_2\Xi_3X_1X_2K_2),
\end{align}
where the following formulations have been used
\begin{align} 
	&\mathbf{a}^r=\boldsymbol{\Gamma} + \mathbf{K} \times X_k  \mathbf{d}^0_k=\boldsymbol{\Gamma} - X_2 K_3 \mathbf{d}^0_1 + X_1 K_3 \mathbf{d}^0_2 + (X_2 K_1+X_1K_2)\mathbf{d}^0_3,\\
	& \Gamma_i = \boldsymbol{\Gamma} \cdot \mathbf{d}_i^0, \quad K_i= \mathbf{K}\cdot\mathbf{d}^0_i, \quad  \Xi_i = \boldsymbol{\Xi} \cdot \mathbf{d}_i^0, \quad \Theta_i = \boldsymbol{\Theta} \cdot \mathbf{d}_i^0,\\
	&(\mathbf{K} \times X_k  \mathbf{d}^0_k) \cdot \mathbf{d}^0_1 = - X_2 \mathbf{K}\cdot\mathbf{d}^0_3,\\
	&(\mathbf{K} \times X_k  \mathbf{d}^0_k) \cdot \mathbf{d}^0_2 =  X_1 \mathbf{K}\cdot\mathbf{d}^0_3,\\
	&(\mathbf{K} \times X_k  \mathbf{d}^0_k)\cdot \mathbf{d}^0_3 = X_2 \mathbf{K}\cdot\mathbf{d}^0_1-X_1 \mathbf{K}\cdot\mathbf{d}^0_2,\\
	&\boldsymbol\Theta\cdot\mathbf{X} =\Theta_1X_1+\Theta_2X_2,\\
	& \mathbf{E}^e = \Xi_1 \mathbf{d}^0_1 +  \Xi_2 \mathbf{d}^0_2 + (\Xi_3+ \Theta_1X^1 + \Theta_2X^2) \mathbf{d}^0_3,\\
	& \mathbf{C}^{\rm linear}=\mathbf{I}+2(\mathbf{a}^r \otimes \mathbf{d}^0_3)^{\rm sym}.
\end{align}

\clearpage
\small

\bibliographystyle{plainnat}
\bibliography{literature}

\end{document}